\documentclass[usenatbib]{mnras}
\usepackage{natbib,amssymb,amsmath,times,graphicx,url,aas_macros}
\usepackage{setspace}      
\usepackage{epstopdf}       
\usepackage{verbatim}       
\usepackage{bm}		  
\usepackage{fleqn}		  
\usepackage{soul}               

\newcommand\sm{$\sim$}
\newcommand\tff{$t_\text{ff}$}
\newcommand\persec{s$^{-1}$}
\newcommand\gpercc{g~cm$^{-3}$}

\newcommand\zetacr{$\zeta_\text{cr}$}
\newcommand\zetam[1]{$\zeta_{#1}$}  
\newcommand\zetaeq[1]{$\zeta_\text{cr}=10^{#1}$ s$^{-1}$}  
\newcommand\zetaapprox[1]{$\zeta_\text{cr}\approx10^{#1}$ s$^{-1}$}  
\newcommand\zetagt[1]{$\zeta_\text{cr} > 10^{#1}$ s$^{-1}$}  
\newcommand\zetage[1]{$\zeta_\text{cr} \gtrsim 10^{#1}$ s$^{-1}$}  
\newcommand\zetale[1]{$\zeta_\text{cr} \lesssim 10^{#1}$ s$^{-1}$}

\newcommand\graina[1]{$a_0 = {#1} \mu$m} 

\newcommand\rhoeq[1]{$\rho_\text{max} = 10^{#1}$ g~cm$^{-3}$}  
\newcommand\rhoapprox[1]{$\rho_\text{max} \approx 10^{#1}$ g~cm$^{-3}$}  
  
\newcommand\rhole[1]{$\rho_\text{max} \lesssim 10^{#1}$ g~cm$^{-3}$}  
\newcommand\rhoge[1]{$\rho_\text{max} \gtrsim 10^{#1}$ g~cm$^{-3}$}  
\newcommand\rhogt[1]{$\rho_\text{max} > 10^{#1}$ g~cm$^{-3}$}  

\newcommand\rhomax{$\rho_\text{max}$}  
  
\newcommand\rhotwoapprox[2]{$\rho_\text{max} \approx #1 \times 10^{#2}$ g~cm$^{-3}$}

\newcommand\rhotwoge[2]{$\rho_\text{max} \gtrsim #1 \times 10^{#2}$ g~cm$^{-3}$}

\defcitealias{PriceBate2007}{PB07}
\defcitealias{WPB2016}{WPB16}

\title[Extreme ionisation rates in star formation]{The effect of extreme ionisation rates during the initial collapse of a molecular cloud core}

\author[Wurster, Bate \& Price]{James Wurster$^{1}$\thanks{j.wurster@exeter.ac.uk}, Matthew R. Bate$^{1}$\thanks{mbate@astro.ex.ac.uk}, and Daniel J. Price$^{2}$ \\
$^{1}$School of Physics and Astronomy, University of Exeter, Stocker Rd, Exeter EX4 4QL, UK \\
$^{2}$Monash Centre for Astrophysics and School of Physics and Astronomy, Monash University, Vic 3800, Australia \\
}
\date{Submitted: Revised: Accepted: }
\pagerange{\pageref{firstpage}--\pageref{lastpage}} \pubyear{2018}

\begin{document}
\label{firstpage}
\bibliographystyle{mnras}
\maketitle

\begin{abstract}

What cosmic ray ionisation rate is required such that a non-ideal magnetohydrodynamics (MHD) simulation of a collapsing molecular cloud will follow the same evolutionary path as an ideal MHD simulation or as a purely hydrodynamics simulation?  To investigate this question, we perform three-dimensional smoothed particle non-ideal magnetohydrodynamics simulations of the gravitational collapse of rotating, one solar mass, magnetised molecular cloud cores, that include Ohmic resistivity, ambipolar diffusion, and the Hall effect.  We assume a uniform grain size of $a_\text{g} = 0.1\mu$m, and our free parameter is the cosmic ray ionisation rate, $\zeta_\text{cr}$.  We evolve our models, where possible, until they have produced a first hydrostatic core.  Models with $\zeta_\text{cr}\gtrsim10^{-13}$ s$^{-1}$ are indistinguishable from ideal MHD models and the evolution of the model with $\zeta_\text{cr}=10^{-14}$ s$^{-1}$ matches the evolution of the ideal MHD model within one per cent when considering maximum density, magnetic energy, and maximum magnetic field strength as a function of time; these results are independent of $a_\text{g}$.  Models with very low ionisation rates ($\zeta_\text{cr}\lesssim10^{-24}$ s$^{-1}$) are required to approach hydrodynamical collapse, and even lower ionisation rates may be required for larger $a_\text{g}$.  Thus, it is possible to reproduce ideal MHD and purely hydrodynamical collapses using non-ideal MHD given an appropriate cosmic ray ionisation rate.  However, realistic cosmic ray ionisation rates approach neither limit, thus non-ideal MHD cannot be neglected in star formation simulations.

\end{abstract}


\section{Introduction}
\label{intro}

Molecular clouds contain magnetic fields \citep[e.g.][]{Crutcher1999,BourkeMyersRobinsonHyland2001,HeilesCrutcher2005,TrolandCrutcher2008} with low ionisation fractions \citep{MestelSpitzer1956} as low as $n_\text{e}/ n_{\text{H}_{2}} = 10^{-14}$ in dense cores \citep{NakanoUmebayashi1986,UmebayashiNakano1990}.   Prior to star formation, ionisation is mostly driven by cosmic rays interacting with the gas and dust, with contributions from radionuclide decay.  After a protostar forms, the protostar itself is thermally ionised and ionises its immediate environment through X-rays.  The ionisation rate depends on the source, with typical rates for cosmic rays, radionuclide decay and X-rays given by $\zeta_\text{cr} \approx 10^{-17}$ s$^{-1} \exp\left(-\Sigma/\Sigma_\text{cr}\right)$ \citep{SpitzerTomasko1968,UmebayashiNakano1981},  $\zeta_\text{r} \approx 7.6\times 10^{-19}$ s$^{-1}$  \citep{UmebayashiNakano2009} and  $\zeta_\text{Xr} \approx 9.6\times 10^{-17}$ s$^{-1} \exp\left(-\Sigma/\Sigma_\text{Xr}\right)$  \citep[e.g.][]{IgeaGlassgold1999,TurnerSano2008}, respectively, where $\Sigma$ is the surface density of the gas, and $\Sigma_\text{cr}$ and $\Sigma_\text{Xr}$ are the characteristic attenuation depths of cosmic rays and X-rays, respectively. 

A completely ionised medium is well represented by ideal magnetohydrodynamics (MHD), while a completely unionised fluid embedded in a magnetic field should be well represented by pure hydrodynamics.  In a partially ionised medium, non-ideal MHD is required, where the three non-ideal effects are electron-ion/neutral drift (Ohmic resistivity), ion-electron drift (Hall effect) and ion-neutral drift (ambipolar diffusion).  Their relative importance depends, amongst other things, on the gas density, number density of charged species (including grains), gas temperature, and magnetic field strength \citep[e.g.][]{WardleNg1999,NakanoNishiUmebayashi2002,TassisMouschovias2007a,Wardle2007,PandeyWardle2008,KeithWardle2014}.  The Hall effect also depends on the direction of the magnetic field with respect to the rotation axis \citep[e.g.][]{BraidingWardle2012sf,BraidingWardle2012accretion,TsukamotoEtAl2015_hall,WPB2016,TsukamotoEtAl2017}.

Many studies have modelled the collapse of a molecular cloud to the first or second Larson core \citep{Larson1969} using non-ideal MHD \citep[e.g.][]{NakanoUmebayashi1986b,FiedlerMouschovias1993,CiolekMouschovias1994,LiShu1996,Mouschovias1996,MouschoviasCiolek1999,ShuGalliLizanoCai2006,MellonLi2009,DuffinPudritz2009,DappBasu2010,MachidaInutsukaMatsumoto2011,LiKrasnopolskyShang2011,KrasnopolskyLiShangZhao2012,DappBasuKunz2012,TomidaEtAl2013,TomidaOkuzumiMachida2015,TsukamotoEtAl2015_oa,TsukamotoEtAl2015_hall,WPB2016,TsukamotoEtAl2017}.  For efficiency, these studies typically assumed that the dominant ionisation source at early times was cosmic ray ionisation and that there was no attenuation.  Thus, the canonically used cosmic ray ionisation rate is \zetaeq{-17}.

The first three-dimensional models of collapsing magnetised molecular clouds were performed using ideal MHD \citep[e.g.][]{PriceBate2007,HennebelleFromang2008,DuffinPudritz2009,HennebelleCiardi2009,CommerconEtAl2010,SeifriedEtAl2011}, despite ideal MHD being a poor approximation to observed molecular cloud environment.  However, these studies provided useful insight into the behaviour of magnetic fields and provided important benchmarks for future studies.  These models were effectively fully ionised, thus what cosmic ray ionisation rate would be required to reproduce these results, assuming non-ideal effects MHD were included?

As a rotating molecular cloud collapses, a dense disc forms \citep[e.g.][]{Larson1972,Tscharnuter1987}.  Given a realistic cosmic ray attenuation rate, the centre of the dense disc should be very weakly ionised or completely neutral, forming a magnetic dead zone \citep{Gammie1996}.  A very weakly ionised medium can be self-consistently modelled with non-ideal MHD, however this can be very expensive to run.  Thus, at what ionisation rate can a medium be treated as purely hydrodynamical?  

The goal of this study is to model the early collapse of a rotating, magnetised molecular cloud core using non-ideal MHD to determine at what cosmic ray ionisation rates (if any) a purely hydrodynamical or an ideal MHD collapse can be recovered.  The free parameter is the cosmic ray ionisation rate, \zetacr, which we held constant throughout each simulation.  Due to the computational expense when low ionisation rates are used, we only model the collapse up to the formation the first hydrostatic core, except in our two lowest ionisation rate models, which never evolved out of the isothermal collapse phase.  In \citet{WursterBatePrice2018_stellar}, we examined how the collapse to stellar core formation changes if one assumes cosmic ray ionisation rates \zetage{-16}.

This paper is organised as follows: In Section~\ref{sec:numerics}, we present our numerical methods, and in Section~\ref{sec:ic} we present our initial conditions and discuss how the initial environment is affected by different cosmic ray ionisation rates.  In Section~\ref{sec:results} we present and discuss our results, and we conclude in Section~\ref{sec:conclusion}.


\section{Numerical method}
\label{sec:numerics}
\subsection{Non-ideal magnetohydrodynamics}
\label{sec:numerics:nimhd}

We solve the equations of self-gravitating, non-ideal magnetohydrodynamics given by
\begin{eqnarray}
\frac{{\rm d}\rho}{{\rm d}t} & = & -\rho (\nabla\cdot \bm{v}), \label{eq:cty} \\
\frac{{\rm d} \bm{v}}{\rm{d} t} & = & -\frac{1}{\rho}\bm{\nabla} \left(P\mathbb{I}\right) - \frac{1}{\rho}\bm{\nabla} \left(\frac{B^2}{2}\mathbb{I} - \bm{B}\bm{B}\right) - \nabla\Phi, \label{eq:mom} \\
\frac{{\rm d} \bm{B}}{\text{d} t} & = & \left(\bm{B} \cdot \nabla\right) \bm{v} - \bm{B} \left(\nabla\cdot \bm{v}\right) + \left.\frac{\text{d} \bm{B}}{\text{d} t}\right|_\text{non-ideal} + \left.\frac{\text{d} \bm{B}}{\text{d} t}\right|_\text{artificial} \label{eq:ind}, \\
\nabla^{2}\Phi & = & 4\pi G\rho, \label{eq:grav}
\end{eqnarray}
where $\text{d}/\text{d}t \equiv \partial/\partial t + \bm{v}\cdot\bm{\nabla}$ is the Lagrangian derivative,  $\rho$ is the density, ${\bm  v}$ is the velocity, $P$ the hydrodynamic pressure, ${\bm B}$ is the magnetic field (which has been normalised such that the Alfv{\'e}n velocity is defined as $v_\text{A}\equiv B/\sqrt{\rho}$ in code units; see \citealt{PriceMonaghan2004}), $\Phi$ is the gravitational potential, $G$ is the gravitational constant, and $\mathbb{I}$ is the identity matrix.  The equation set is closed by a barotropic equation of state,
\begin{equation}
\label{eq:eos}
P = \left\{ \begin{array}{l l} c_\text{s,0}^2\rho; 		    &  \rho < \rho_\text{c}, \\
                                           c_\text{s,0}^2\rho_\text{c}\left(\rho             /\rho_\text{c}\right)^{7/5};      &  \rho_\text{c} \leq \rho < \rho_\text{d}, \\
                                           c_\text{s,0}^2\rho_\text{c}\left(\rho_\text{d}/\rho_\text{c}\right)^{7/5} \left(\rho/\rho_\text{d}\right)^{11/10};      &  \rho \geq \rho_\text{d},
\end{array}\right.
\end{equation}
where $c_\text{s,0}$ is the initial isothermal sound speed, and the density thresholds are $\rho_\text{c} = 10^{-14}$ \gpercc \ and $\rho_\text{d}~=~10^{-10}$ g~cm$^{-3}$.

The non-ideal MHD term in \eqref{eq:ind} is
\begin{eqnarray}
\left.\frac{\text{d} \bm{B}}{\text{d} t}\right|_\text{non-ideal}= &-&\bm{\nabla} \times \left[  \eta_\text{OR}      \left(\bm{\nabla}\times\bm{B}\right)\right]                                                                   \label{eq:ohm} \\
                                                                                             &-&\bm{\nabla} \times \left[  \eta_\text{HE}       \left(\bm{\nabla}\times\bm{B}\right)\times\bm{\hat{B}}\right]                                      \label{eq:hall} \\
                                                                                             &+&  \bm{\nabla} \times \left\{ \eta_\text{AD}\left[\left(\bm{\nabla}\times\bm{B}\right)\times\bm{\hat{B}}\right]\times\bm{\hat{B}}\right\},  \label{eq:ambi}
\end{eqnarray}
where the non-ideal coefficients for Ohmic resistivity, the Hall effect, and ambipolar diffusion terms are given in (e.g.) \citet{WardleNg1999,Wardle2007}.

To calculate the number densities of the charged species and thus the non-ideal MHD coefficients, we use Version 1.2.1 of the \textsc{Nicil} library \citep{Wurster2016}.  The maximum temperature reached in this study will be $T < 500$~K, thus cosmic rays will be the only ionisation source, since we intentionally ignore ionisation from radionuclide decay in order to test the effect of low ionisation rates.  Cosmic rays can create two species of negatively charged ions: a light ion species based upon hydrogen and helium components and a heavy ion species with the mass of magnesium \citep[e.g.][]{AsplundEtAl2009}.  We include three species of grains which can absorb free electrons to become negatively charged $n_\text{g}^-$, or lose electrons through collisions to become positively charged $n_\text{g}^+$, or remain neutral $n_\text{g}^0$.  The total number density of grains is dependent on the local gas density, viz.,
\begin{equation}
n_\text{g} = f_\text{fg}\frac{m_\text{n}}{m_\text{g}}n_\text{gas}
\end{equation}
where $f_\text{dg}$ is the gas to dust ratio, $m_\text{n}$ and $m_\text{g}$ are the masses of a neutral particle and dust grain, respectively, and $n_\text{gas}$ is the gas number density.  To conserve gain number density, $n_\text{g} = n_\text{g}^- +n_\text{g}^0 + n_\text{g}^+$.

\subsection{Smoothed particle non-ideal magnetohydrodynamics}
\label{sec:numerics:spnimhd}
Our calculations are carried out using the 3D smoothed particle magnetohydrodynamics (SPMHD) code \textsc{Phantom} \citep{Phantom2017all} with the inclusion of self-gravity and non-ideal MHD \citep{Wurster2016}.  The density of each SPH particle $a$ is calculated by iteratively solving
\begin{equation}
\label{eq:density}
\rho_a  = \sum_b m_b W_{ab}(h_a); \hspace{1em} h_{a} = h_\text{fac} \left(\frac{m_{a}}{\rho_{a}}\right)^{1/3}
\end{equation}
using the Newton-Raphson method, where we sum over all neighbours $b$, $m_a$ and $h_a$ are the particle's mass and smoothing length, respectively, $W_{ab}$ is the smoothing kernel, and $h_\text{fac}=1.2$ is a coefficient required to obtain \sm58 neighbours when using the adopted cubic spline kernel.

The remainder of the discretised SPMHD equations are readily available in the literature (e.g. see review by \citealp{Price2012}), and we use the same form as given in \citet{WPB2016}.  We enforce the divergence-free condition on the magnetic field using the constrained hyperbolic/parabolic divergence cleaning algorithm described in \citet{TriccoPrice2012} and \citet*{TriccoPriceBate2016}.

In ideal MHD, artificial resistivity is required for magnetic stability (i.e. the final term in Eqn.~\ref{eq:ind}); as per convention, artificial resistivity is included in all of our simulations both for consistency and for the possibility that physical and artificial resistivity are important in different regions.  We use the form given by \citet{PriceMonaghan2004,PriceMonaghan2005}, however, the signal velocity is instead given by $v_{\text{sig},ab} = \left| \bm{v}_{ab} \times \hat{\bm{r}}_{ab}\right|$ \citep{Phantom2017all}.  A comparison of artificial resistivity algorithms presented in \citet{WursterBatePriceTricco2017} showed that the method used here is the least dissipative of all SPMHD algorithms used to date, especially during the collapse to form the first core. Hence, our results are dominated by physical and not artificial resistivity.

\subsection{Timestepping}
\label{sec:numerics:dt}
In ideal MHD, the limiting timestep for particle $a$ is typically the Courant-Friedrichs-Lewy (CFL) condition,
\begin{equation}
\label{eq:dt:cfl}
\text{d}t_{\text{CFL},a} = \frac{C_\text{CFL}h_a}{\sqrt{c_{\text{s},a}^2 + v_{\text{A},a}^2}},
\end{equation}
where $C_\text{CFL} = 0.3 < 1.0$ is the dimensionless Courant number and $c_{\text{s},a}$ is the sound speed.  However, non-ideal effects each add a new time-constraint, viz.,
\begin{equation}
\label{eq:dt:ni}
\text{d}t_{\text{OR},a} = \frac{C_\text{NI}h_a^2}{\eta_{\text{OR},a}}, \hspace{1em} \text{d}t_{\text{HE},a} = \frac{C_\text{NI}h_a^2}{\left|\eta_{\text{HE},a}\right|},  \hspace{1em}\text{d}t_{\text{AD},a} = \frac{C_\text{NI}h_a^2}{\eta_{\text{AD},a}}, 
\end{equation}
where $C_\text{NI} = 1/2\pi$ is a  dimensionless coefficient analogous to the Courant number.  Test cases with ambipolar diffusion show that the non-ideal MHD timestep can be \sm40-50 shorter than the CFL timestep \citep[e.g.][]{MaclowNormanKoniglWardle1995,WPA2014}, however, in realistic problems, the minimum non-ideal MHD timestep can be several hundred times shorter in quickly evolving, dense regions.

Super-timestepping \citep{AlexiadesAmiezGremaud1996} is used to relax the conditions given by \eqref{eq:dt:ni} for the diffusive terms that are parabolic in nature (i.e. Ohmic resistivity and ambipolar diffusion).  This involves taking $N_\text{sts} < N_\text{real} \approx \text{d}t_{\text{CFL}} /\min\left(\text{d}t_\text{OR},\text{d}t_\text{AD}\right)$ steps of d$\tau_j$ where $j=1..N_\text{sts}$ and $\text{d}\tau_j > \text{d}\tau_{j+1}$ and requiring stability only at the end of $N_\text{sts}$ steps rather than at the end of every step.  The best possible speed-up yields \citep[e.g.][]{ChoiKimWiita2009},
\begin{equation}
\label{eq:nsts}
N_\text{sts} = \text{int}\left[\sqrt{\frac{\text{d}t_\text{CFL}}{k \cdot \min\left(\text{d}t_\text{OR},\text{d}t_\text{AD}\right)}}\right] + 1
\end{equation}
where we set $k=0.9$.  For added stability, we first sub-divide $\text{d}t_{\text{CFL}}$ by positive integer $n$ such that $20 \text{d}\tau_1 \gtrsim \text{d}t_{\text{CFL}}/n$ and then take $nN_\text{sts} < N_\text{real}$ steps per $\text{d}t_{\text{CFL}}$; this subdivision by $n$ is only required in extreme environments where multiple physical processes are simultaneously contributing to a complex evolution, or for very low ionisation rates (i.e. \zetale{-24}).

Given the hyperbolic nature of the Hall effect, we are required to solve this timestep explicitly.  Methods \citep[e.g.][]{OsullivanDownes2007,MeyerBalsaraAslam2012} have been proposed to sub-step with the Hall term, but these have yet to be implemented into \textsc{Phantom}.

\section{Initial conditions}
\label{sec:ic} 

Our models are similar to those used in our previous studies (\citealp{PriceBate2007,BateTriccoPrice2014,WPB2016}) and consists of a spherical cloud of radius $R=4\times~10^{16}$~cm = 0.013~pc and density $\rho_0=7.43\times 10^{-18}$ \gpercc \ which is placed inside a low-density box of edge length $l = 4R$ and a density contrast of 30:1; the cloud and surrounding medium are in pressure equilibrium.  This allows the cloud to be modelled self-consistently, and we use quasi-periodic boundary conditions at the edge of the box, in which SPH particles interact magnetohydrodynamically `across the box', but not gravitationally.   Our simulations use $10^6$ particles in the sphere which are initialised on a regular close-packed lattice.
 
The initial cloud has mass $M=1$~M$_{\odot}$, rotational velocity $\Omega = 1.77\times 10^{-13}$ rad s$^{-1}$, and sound speed $c_\text{s,0} = 2.19\times 10^4$~cm~s$^{-1}$ (i.e. $T_0 = 13.5$~K).  We thread the cloud with a uniform magnetic field that is anti-aligned with the axis of rotation, i.e. $\bm{B} = -B_0\hat{\bm{z}}$, which will promote disc formation in the presence of the Hall effect \citep[e.g.][]{BraidingWardle2012sf,BraidingWardle2012accretion,TsukamotoEtAl2015_hall,WPB2016}; this configuration yields $\left(\bm{\nabla}\times \bm{B}_0\right)_0 = 0$.  The magnetic field has an initial strength of $B_0 = 1.63 \times 10^{-4}$ G, which corresponds to a normalised mass-to-flux ratio of $\mu_0 \equiv \left(M/\Phi_\text{B}\right)_0/\left(M/\Phi_\text{B}\right)_\text{crit} = 5$, where $\left(M/\Phi_\text{B}\right)_0 \equiv M/\left(\pi R^2 B\right)$ is the initial mass-to-flux ratio and $\left(M/\Phi_\text{B}\right)_\text{crit} = c_1/\left(3\pi\right)\sqrt{5/G }$ is the critical value where magnetic fields prevent gravitational collapse altogether; $M$ is the total mass contained within the cloud, $\Phi_\text{B}$ is the magnetic flux threading the surface of the (spherical) cloud at radius $R$ assuming a uniform magnetic field of strength $B$, $G$ is the gravitational constant and $c_1 \simeq 0.53$ is a parameter numerically determined by \citet{MouschoviasSpitzer1976}.  The free-fall time is $t_\text{ff}=2.4\times~10^4$~yr, which is the characteristic timescale for this study.  

The non-ideal MHD models use the default values included in the \textsc{Nicil} library \citep{Wurster2016}.  The dust grains have a radius and bulk density of $a_{\rm g} = 0.1 \mu$m and $\rho_{\rm b} = 3$ g~cm$^{-3}$ \citep{PollackEtAl1994}, respectively, and the dust-to-gas ratio is $f_\text{dg} = 0.01$.  The mass of the neutral particle is based upon the hydrogen and helium abundance, thus $m_\text{n} = 2.31$m$_\text{p}$, where m$_\text{p}$ is the mass of a proton.  We test 15 cosmic ray ionisation rates in the range $\zeta_\text{cr} \in \left[10^{-30},10^{-10}\right]$ \persec, which are indicated by the vertical lines in Fig.~\ref{fig:ic:nicil}.  Our models will be named after their ionisation rate such that model $\zeta_X$ has a cosmic ray ionisation rate of \zetaeq{-X}.  The ideal MHD and purely hydrodynamical models will be referred to as iMHD and HD, respectively.

Since the time constraint imposed by non-ideal MHD quickly becomes prohibitively expensive for low ionisation rates, our focus is on comparing the very early phases of the collapses, typically just prior to or just after entering the first hydrostatic core phase, which begins at \rhoapprox{-13}.  The maximum density analysed in this study is \rhoapprox{-11}, however, several of the low ionisation rate models end at \rhole{-14}.

\subsection{Initial behaviour of the charged species and non-ideal MHD coefficients}
\label{sec:IC:nimhd}

\begin{figure}
\begin{center}
\includegraphics[width=\columnwidth]{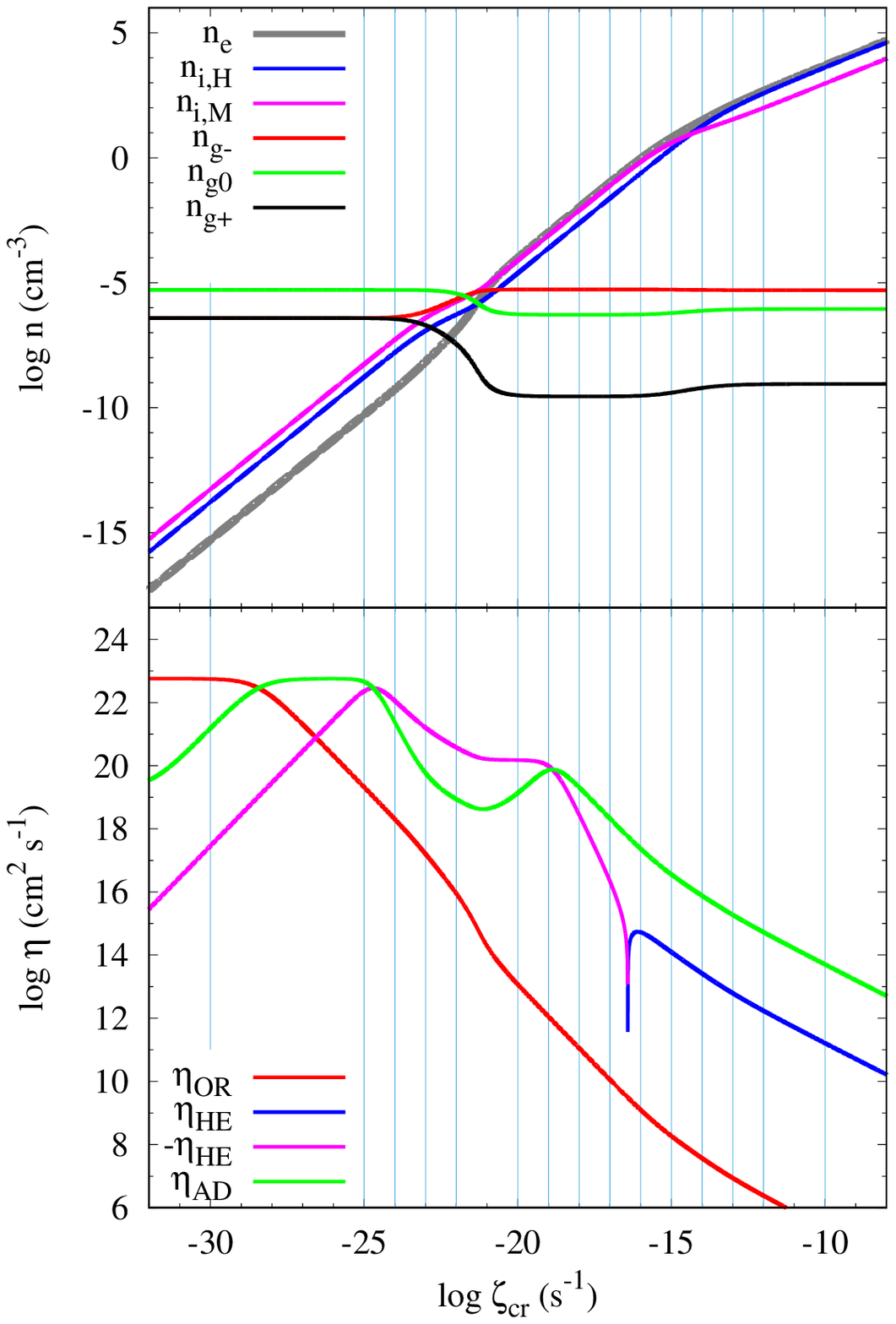}
\caption{The species number densities (top) and non-ideal MHD coefficients $\eta$ (bottom) calculated using our initial conditions of $\rho_0=7.43\times~10^{-18}$ \gpercc, $B_0 = 1.63 \times 10^{-4}$ G (i.e. $\mu_0 = 5$) and $c_\text{s,0} = 2.19\times 10^4$~cm~s$^{-1}$ (i.e. $T_0 = $13.5~K).  The vertical lines represent the values of the cosmic ray ionisation rate, \zetacr, that are included in our suite.  The grain populations are the dominant species at low ionisation rates, and the positively charged ions and electrons are dominant at high ionisation rates.  This turnover is reflected in $\eta$, where $\eta_\text{AD} < |\eta_\text{HE}|$.  Increasing \zetacr \ does not lead to a monotonic change in $\eta$, thus models with different initial values of \zetacr \ will start with different non-ideal effects controlling the evolution.}
\label{fig:ic:nicil}
\end{center}
\end{figure}

Fig.~\ref{fig:ic:nicil} shows the species number densities and the non-ideal MHD coefficients, calculated using our initial conditions.  
At constant density, temperature and magnetic field strength, the number densities and non-ideal MHD coefficients are strongly dependent on the cosmic ionisation rate, \zetacr.

At ionisation rates of \zetale{-24}, cosmic rays are unable to ionise ions rapidly enough for the ions and electrons to significantly contribute to the charged species populations; at these rates, the charged species are from grain collisions that transfer electrons to make a positively and negatively charged grain population, with $n_\text{g}^- \approx n_\text{g}^+$.  At ionisation rates of \zetage{-20}, the ion and electron populations are several orders of magnitude more populous than the charged grain number densities.  However, the grains have a much larger mass than the ions (i.e. $m_\text{g} = 7.5\times 10^9$m$_\text{p}$ compared to $m_\text{light ion} = 2.31 $m$_\text{p}$ and $m_\text{heavy ion} = 24.3 $m$_\text{p}$), thus still contribute non-trivially to the value of the non-ideal MHD coefficients even at very high ionisation rates.

For \zetage{-20} (recall that the canonical cosmic ionisation rate is \zetaeq{-17}), the number density of ions and electrons is similar, and the ionisation fraction reaches $n_\text{e,i}/(n_\text{n}+n_\text{i}) \approx 0.003$ at \zetaapprox{-10}; thus, even at high cosmic ray ionisation rates, thermal ionisation or another source is required to fully ionise the medium. 

Using our given initial conditions, the initial non-ideal MHD coefficients have six regimes:
\begin{enumerate}
\item \ \ \ \ \ \  \ \ \ \ \ \ \ \ \ \ \ \ \ \ \ \  $ \zeta_\text{cr}/\text{s}^{-1} \lesssim 4\cdot 10^{-29}$: $\eta_\text{OH} > \eta_\text{AD} > -\eta_\text{HE}$;
\item $4\cdot 10^{-29} \lesssim \zeta_\text{cr}/\text{s}^{-1} \lesssim 3\cdot 10^{-27}$: $\eta_\text{AD} > \eta_\text{OR} > -\eta_\text{HE}$;
\item $3\cdot 10^{-27} \lesssim \zeta_\text{cr}/\text{s}^{-1} \lesssim 2\cdot 10^{-25}$: $\eta_\text{AD} > -\eta_\text{HE} > \eta_\text{OR}$;
\item $2\cdot 10^{-25} \lesssim \zeta_\text{cr}/\text{s}^{-1} \lesssim 1\cdot 10^{-19}$: $-\eta_\text{HE} > \eta_\text{AD} > \eta_\text{OR}$;
\item $1\cdot 10^{-19} \lesssim \zeta_\text{cr}/\text{s}^{-1} \lesssim 4\cdot 10^{-17}$: $\eta_\text{AD} > -\eta_\text{HE} > \eta_\text{OR}$;
\item $4\cdot 10^{-17} \lesssim \zeta_\text{cr}/\text{s}^{-1}$:  \ \ \ \ \ \  \ \ \ \ \ \ \ \ \ \ \ \ \ \  $\eta_\text{AD} > \eta_\text{HE} > \eta_\text{OR}$.
\end{enumerate}

Although ambipolar diffusion is typically the dominant effect, the Hall effect is the dominant term in region (iv), which is the same region where grains transition from higher number densities compared to the ions to lower number densities.  At very low ionisation rates, the Ohmic coefficient is approximately constant, since it is dependent on the Ohmic conductivity $\sigma_\text{O}$, which is approximately constant due to the grain number densities.  The Hall coefficient rapidly decreases at low ionisation rates due to its dependence on the Hall conductivity $\sigma_\text{H}$, which will rapidly decreases for $n_\text{g}^- \approx n_\text{g}^+ \gg n_\text{i}$.  The ambipolar coefficient is also dependent on the Hall conductivity, but via the perpendicular conductivity, hence its delayed decrease.  At high ionisation rates, all three terms decrease rapidly as the cloud becomes more ionised.  Thus, all three terms have less of an effect on the evolution of the cloud in an absolute sense.

These results are qualitatively similar to those that are obtained from using different $\rho_0$, $B_0$ and $T_0$.  Therefore, the numerical results will necessarily differ if we change our initial conditions, however, our qualitative results will be independent of them.

\subsection{Grain properties}
\label{sec:IC:grain}

Although a uniform grain size is not realistic, they are common in numerical models.  We use the uniform grain size of $a_0 = 0.1\mu$m to match our previous studies \citep{WPB2016,WPB2017,WursterBatePrice2018_stellar} and to agree with the fiducial value suggested by \citet{PollackEtAl1994}.  Uniform grain sizes of smaller radius were used in \citet{TsukamotoEtAl2015_oa,TsukamotoEtAl2015_hall,TsukamotoEtAl2017}.  

\begin{figure}
\begin{center}
\includegraphics[width=0.8\columnwidth]{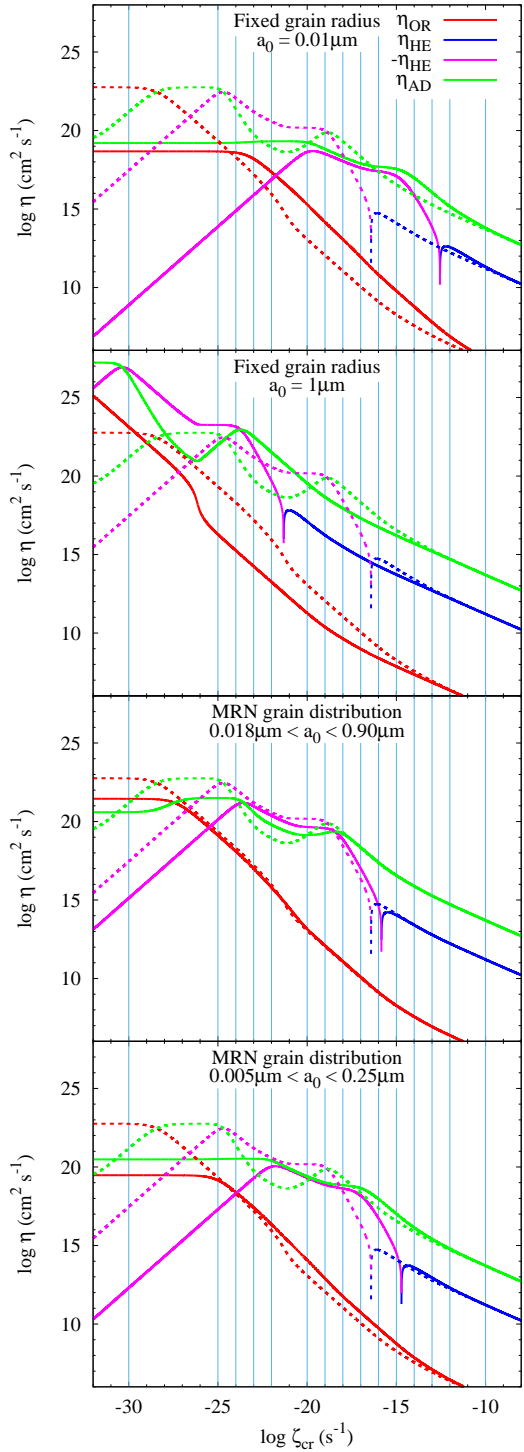}
\caption{The non-ideal MHD coefficients using using our initial conditions as in the bottom panel of Fig.~\ref{fig:ic:nicil}.  The dotted lines in each panel are calculated using our fiducial uniform grain size of \graina{0.1}, and the solid lines are calculated using the grain size/distribution listed in the panel.  The MRN distribution is calculated assuming 40 bins of equal width in log-space.  The coefficients have a greater dependence on grain properties at lower cosmic ray ionisation rates, where the  coefficients can differ by up to 9 dex; at high ionisation rates (\zetage{-13}) the coefficients typically differ by less than 10 per cent.}
\label{fig:ic:nicil:grains}
\end{center}
\end{figure}

An alternative to the uniform grain size is the \citet*{MathisRumplNordsieck1977} (MRN) grain distribution,
\begin{equation}
\label{eqn:mrn}
\frac{\text{d}n_\text{g}(a)}{\text{d}a} = An_\text{H}a^{-3.5},
\end{equation}
where $n_\text{H}$ is the number density of the hydrogen nucleus, $n_\text{g}(a)$ is the number density of grains with a radius smaller than $a$, and $A = 1.5\times10^{-25}$ cm$^{2.5}$ \citep{DraineLee1984}.  In this distribution, there are more grains with smaller radii, thus the smaller grains will more strongly influence the evolution than the larger grains.  Fig.~\ref{fig:ic:nicil:grains} shows the non-ideal MHD coefficients as a function of \zetacr \ using a uniform grain size of  $a_0 = 0.01$ and $1\mu$m (top two panels), and using the MRN grain distribution using the ranges suggested in \citet{KunzMouschovias2009} and \citet{WardleNg1999} (bottom two panels, respectively).

At high ionisation rates (\zetage{-13}), the coefficients differ by less than 10 per cent, except for \graina{0.1} where the 10 per cent agreement is only for \zetage{-10}.  Thus, for high ionisation rates, we expect the grain properties to play a minimal role in the evolution of the system.  As \zetacr \ decreases to realistic rates (\zetaapprox{-17}), the coefficients become more dependent on the grain properties, although the coefficients for our fiducial grain size and the MRN distribution using the \citet{KunzMouschovias2009} range differ by less than a factor of 1.3.

At low ionisation rates (\zetale{-25}), the coefficients can differ by up to nine orders of magnitude.  
For \graina{1}, the coefficients are larger than using our fiducial grain size, suggesting that these systems will approach the hydrodynamical limit at higher ionisation rates due to greater non-ideal MHD effects.  
For \graina{0.01} and the MRN distributions, the coefficients are typically lower than for our fiducial grain size suggesting these models will be slightly more ideal than our fiducial models.  Given that the MRN distributions are the more realistic models, the results we present in Section~\ref{sec:res:hydro} will be upper limits such that using the MRN distribution would require even lower ionisation rates to approximate the hydrodynamical case than models using our fiducial uniform grain size of \graina{0.1}.

\section{Results}
\label{sec:results}

\subsection{Limiting models: Hydrodynamic and Ideal MHD}
\label{sec:res:limiting}

\begin{figure}
\begin{center}
\includegraphics[width=\columnwidth]{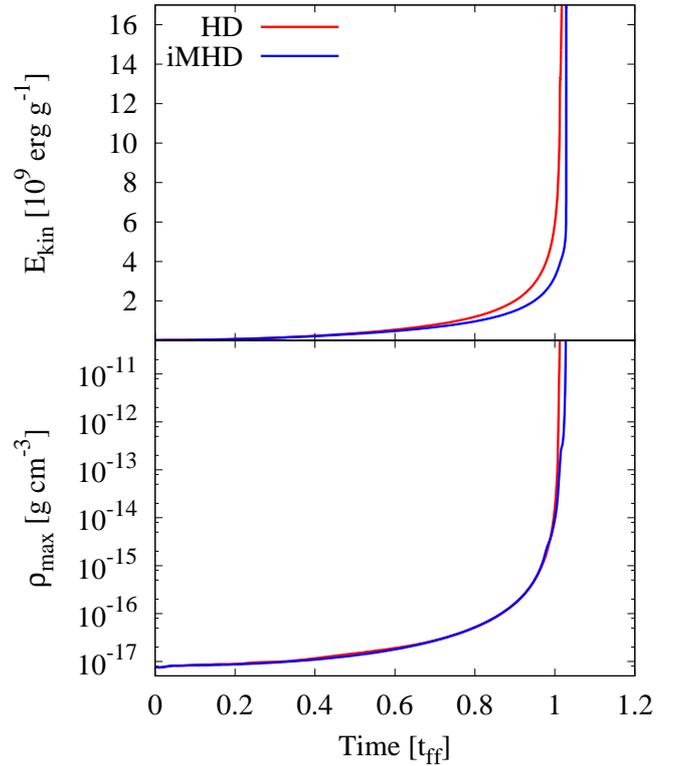}
\caption{The evolution of the total kinetic energy (top) and the maximum density (bottom) for the purely hydrodynamic (HD) and ideal MHD (iMHD) models.  The kinetic energy differs by more than 10 per cent for $t \gtrsim 0.51$\tff, and the maximum density differs by more than 10 per cent at $t \gtrsim 0.99$ \tff.}
\label{fig:res:HI}
\end{center}
\end{figure}

The iMHD and HD models represent the limiting models such that the models are totally ionised and totally neutral, respectively.  Fig.~\ref{fig:res:HI} shows their evolution of the total kinetic energy and the maximum density.  

Qualitatively, both models follow similar trends, with a slow evolution until $t\approx$ \tff, at which time the collapse occurs very rapidly.  The presence of strong magnetic fields delays the collapse of the molecular cloud, such that, by $t \approx 0.99$ \tff, the maximum densities differs by 10 per cent, and it then takes iMHD \sm370 yr longer to reach \rhoapprox{-11} than HD.  

The kinetic energy begins to diverge almost immediately with it differing by 10 per cent after $t \approx 0.51$ \tff, with more kinetic energy in HD than in iMHD at any given time.  This is expected since the magnetic field supports the cloud against gravitational collapse, thus the gas collapses slightly slower.

\subsection{Transition models: Non-ideal MHD}
\begin{figure}
\begin{center}
\includegraphics[width=\columnwidth]{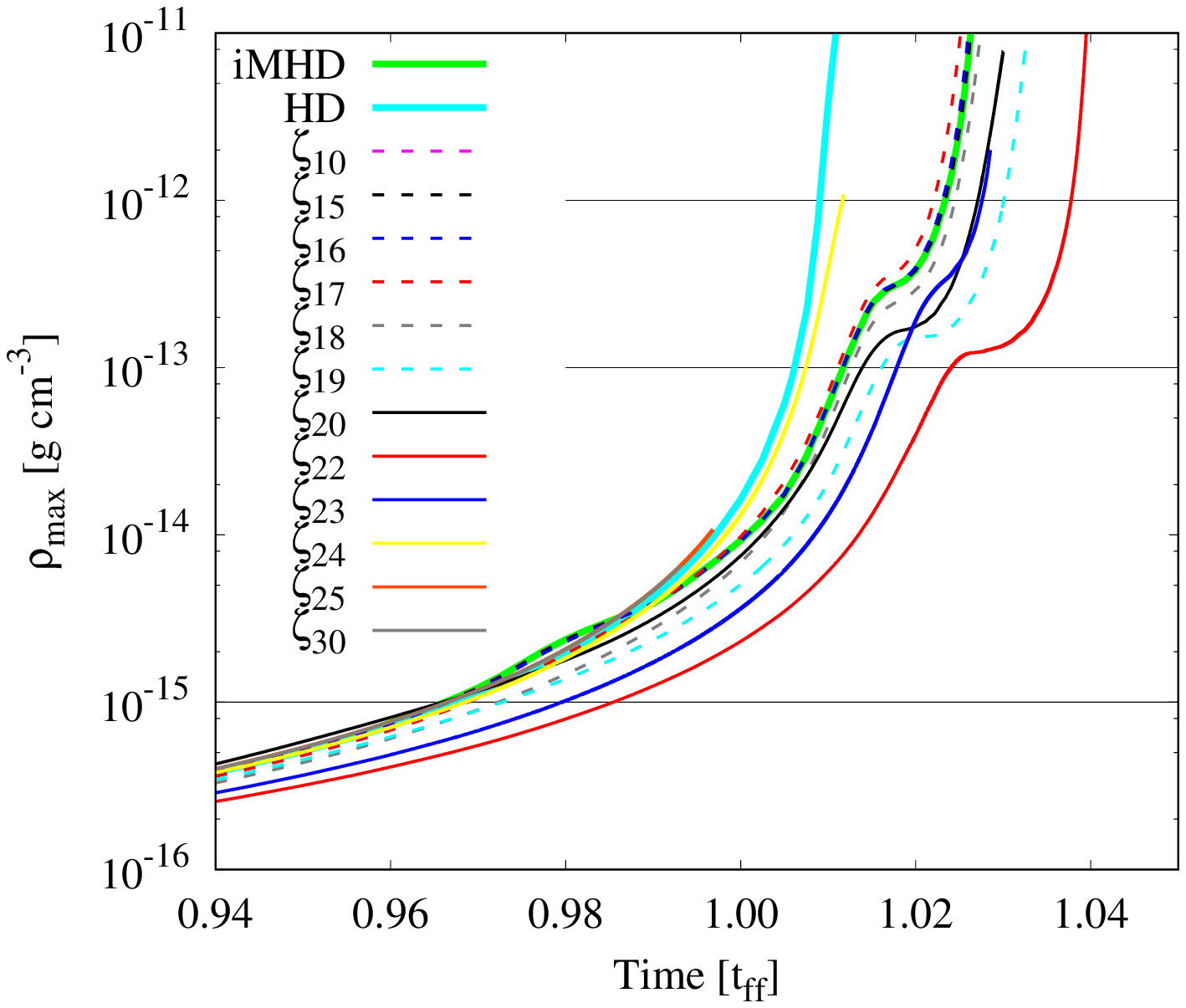}
\caption{The evolution of the maximum density for selected models for $t > 0.94$\tff.  The thick green and cyan lines represent models iMHD and HD, respectively, whose full evolution is shown in the bottom panel of Fig.~\ref{fig:res:HI}.  Models with \zetage{-16} all lie on top of the iMHD curve, thus only two have been shown for clarity.  Model \zetam{17}  evolves slightly faster than iMHD, while \zetam{18} -  \zetam{23} evolve slower.  The low ionisation rate models with \zetale{-24} evolve similarly to HD.  The maximum densities for HD and \zetam{22}-\zetam{30} do not coincide with the centre of the core.  Not all models have reached \rhoeq{-11} due to computational limitations.  The horizontal lines match the maximum densities shown in Figs.~\ref{fig:res:slice:rho}-\ref{fig:res:slice:vy}.}
\label{fig:res:rhoVt}
\end{center}
\end{figure}

\begin{figure*}
\begin{center}
\includegraphics[width=\textwidth]{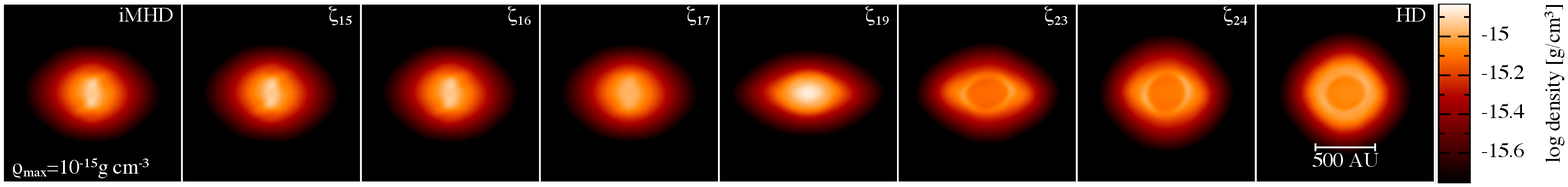}
\includegraphics[width=\textwidth]{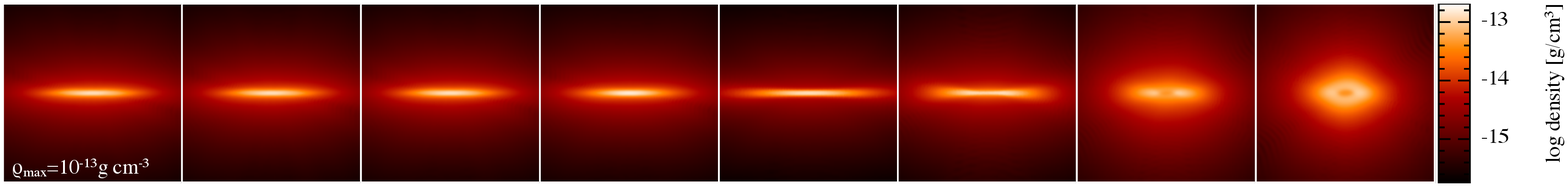}
\includegraphics[width=\textwidth]{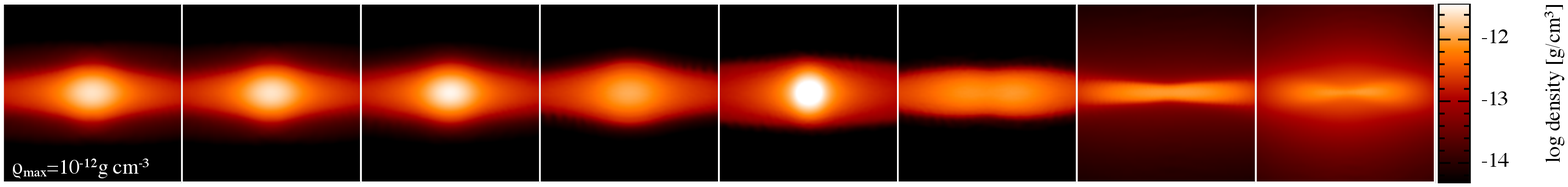}
\caption{Density slice through the core of the clouds of selected models at $\rho_\text{max} \approx 10^{-15}$, $10^{-13}$ and $10^{-12}$ \gpercc.  Frame sizes and colour bar range change with each row to better show the structure at each $\rho_\text{max}$.   At \rhoapprox{-15}, a dense column of gas has formed along the rotation axis in the higher ionisation rate models (\zetage{-17}),  an oblate spheroid has formed for mid-range ionisation rates, and the collapse is approximately spherical for low ionisation rates (\zetale{-23}); the maximum density is not in the core for the models with \zetale{-22}.  At \rhoapprox{-13}, the scale height of the discs decreases from iMHD to \zetam{22} and then increases again.}
\label{fig:res:slice:rho}
\end{center}
\end{figure*}

\begin{figure*}
\begin{center}
\includegraphics[width=0.87\textwidth]{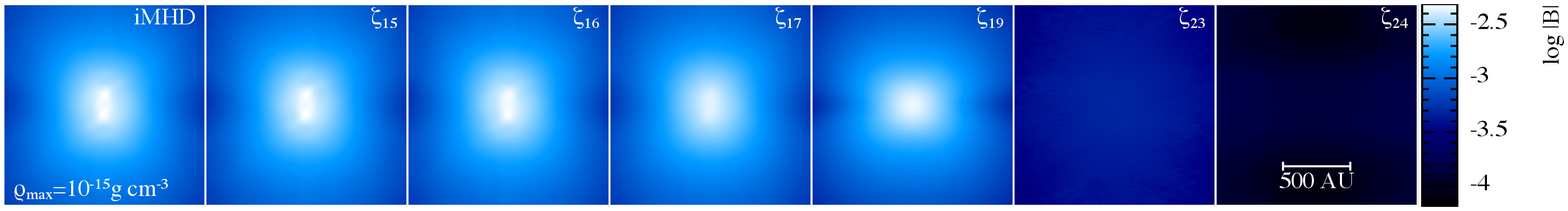}
\includegraphics[width=0.87\textwidth]{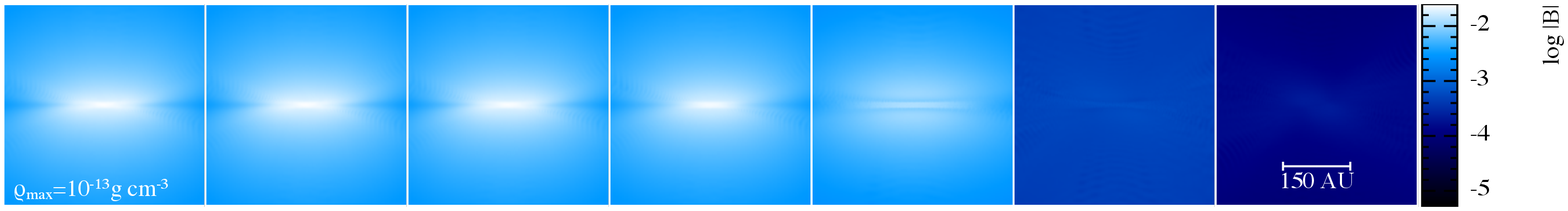}
\includegraphics[width=0.87\textwidth]{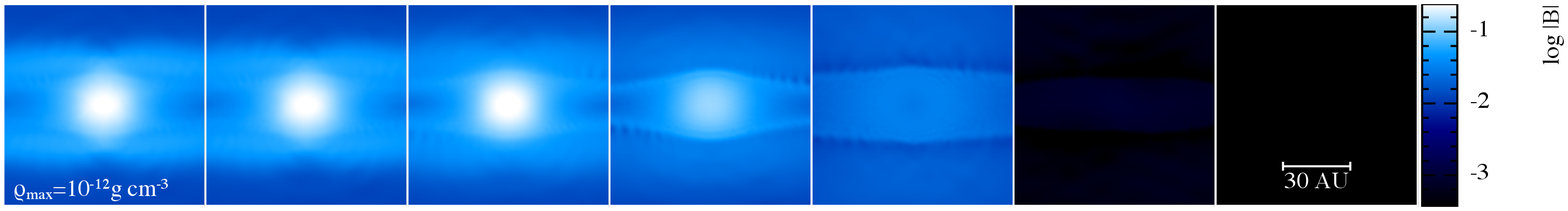}
\caption{Magnetic field strength through the core of the clouds of selected models as in Fig.~\ref{fig:res:slice:rho}.  At \rhoapprox{-15}, the dense column of gas has an enhanced magnetic field strength in the higher ionisation rate models (\zetage{-17}), while the magnetic field strength is approximately uniform at low ionisation rates (\zetale{-23}).  By \rhoapprox{-12}, the models with \zetale{-19} have an unstructured magnetic field that is stronger in the midplane.}
\label{fig:res:slice:rB}
\end{center}
\end{figure*}

\begin{figure*}
\begin{center}
\includegraphics[width=\textwidth]{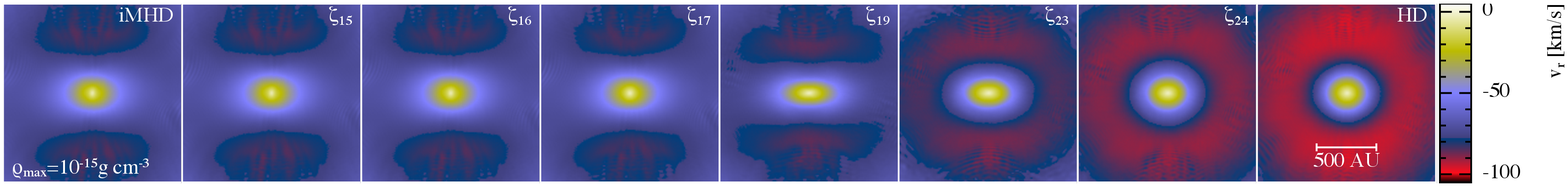}
\includegraphics[width=\textwidth]{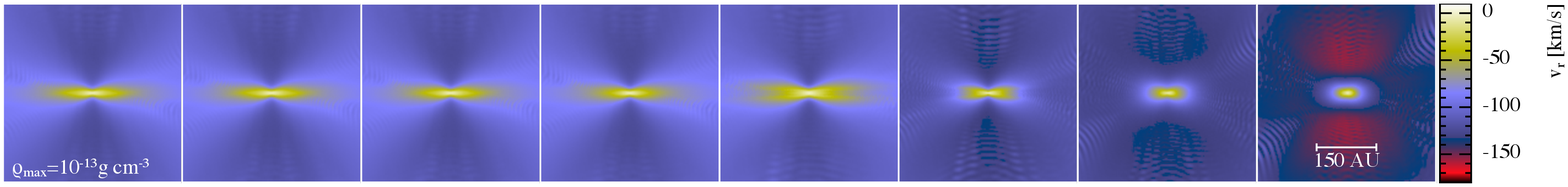}
\includegraphics[width=\textwidth]{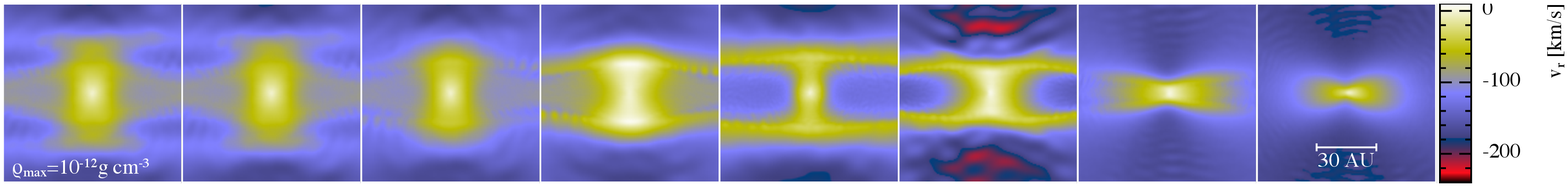}
\caption{Radial velocity slices through the core of the clouds of selected models as in Fig.~\ref{fig:res:slice:rho}.  The initial radial infall (i.e. where $v_\text{r} < 0$) is approximately spherical for the low ionisation rate models (\zetale{-24}) and faster at larger radii from the core than for models with higher rates (\zetagt{-24}).  At larger $\rho_\text{max}$, the infall becomes less spherical in all models, and the infall rate is faster in the midplane for the magnetised models while the vertical infall is faster in HD.}
\label{fig:res:slice:vr}
\end{center}
\end{figure*}

\begin{figure*}
\begin{center}
\includegraphics[width=\textwidth]{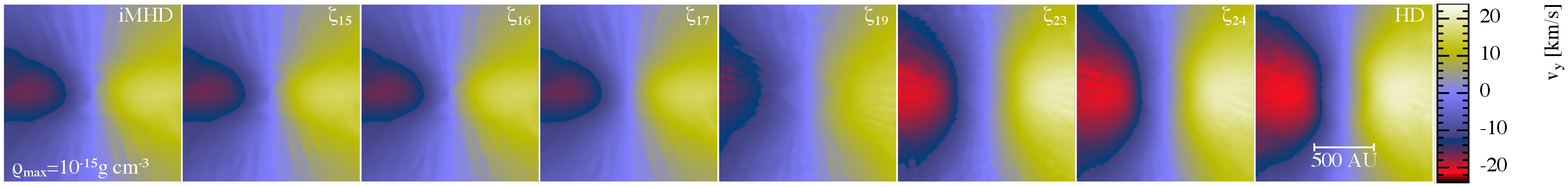}
\includegraphics[width=\textwidth]{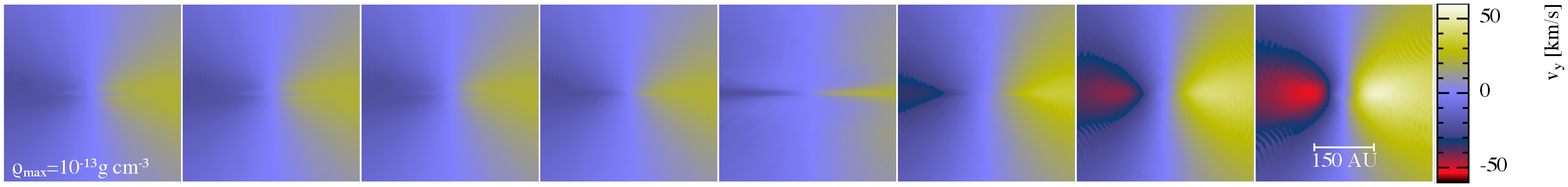}
\includegraphics[width=\textwidth]{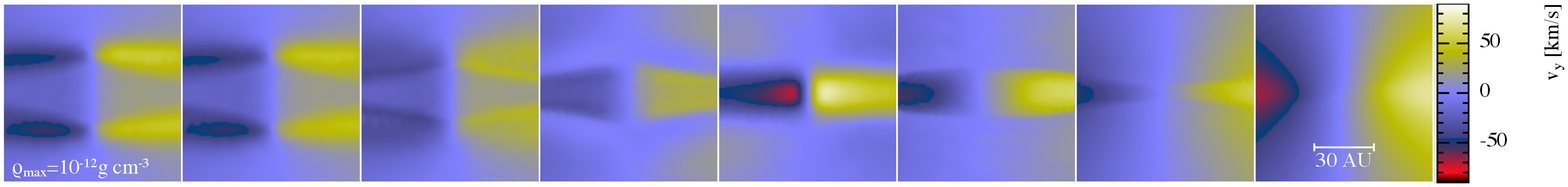}
\caption{Azimuthal velocity slices through the core of the clouds of selected models as in Fig.~\ref{fig:res:slice:rho}.  The rotation speed of the cloud increases as the cosmic ray ionisation rate decreases and as time increases.  At \rhoapprox{-12}, the rotational speed is faster above and below the disc for \zetale{-16}; a counter-rotating envelope is beginning to form in \zetam{18} and \zetam{19}.}
\label{fig:res:slice:vy}
\end{center}
\end{figure*}

We define any model that includes non-ideal MHD as a transition model since the ionisation fraction is neither one (iMHD) nor zero (HD).   Fig.~\ref{fig:res:rhoVt} shows the late evolution ($ t > 0.94$\tff) of the maximum density for selected models, and Figs.~\ref{fig:res:slice:rho}-\ref{fig:res:slice:vy} show cross sections of the density, magnetic field strength, radial and azimuthal velocities of selected models at  $\rho_\text{max} \approx 10^{-15}$, $10^{-13}$ and $10^{-12}$ \gpercc.

The models with \zetage{-16} evolve similarly to iMHD and will be discussed in more detail in Section~\ref{sec:res:ideal} below.  These clouds are magnetically supported, thus form a dense collimated structure with strong magnetic fields, since initially, the gas is free to collapse along the rotation axis.  The models with \zetale{-24} follow a similar evolutionary path as HD and will be discussed in Section~\ref{sec:res:hydro}.  These models follow an initially spherical collapse, however, by  \rhogt{-13}, a thick, rotationally supported disc has formed; throughout their evolution, they retain an approximately uniform magnetic field.

Models \zetam{16} to \zetam{24} do not represent a smooth transition between the evolutionary paths of iMHD and HD as the initial ionisation rate is decreased, with the exception of the magnetic field strength and geometry.  Agreeing with intuition, \zetam{17} evolves similarly to but slightly faster than iMHD, thus is in the region bracketed by iMHD and HD in Fig.~\ref{fig:res:rhoVt}.  As expected, at our selected \rhomax \ snapshots, its central gas distribution is more diffuse and has a weaker magnetic field strength than the models with \zetage{-16}.

The models with $10^{-23} \lesssim \zeta_\text{cr}/\text{s}^{-1} \lesssim 10^{-18}$ evolve slower than iMHD, with \zetam{22} having the longest evolutionary time.  This is a result of the Hall effect.  At these early times, the rotation is beginning to convert the poloidal magnetic field, $|B_\text{p}| = \sqrt{B_\text{r}^2 + B_\text{z}^2}$, into a toroidal magnetic field, $|B_\phi |$.  Since the field lines remain closed, there are both $\pm B_\phi \hat{\bm{\phi}}$ components, and the Hall effect enhances one component and decreases the other.  At this stage, the initial direction of the magnetic field is relatively unimportant for the characteristics that we investigate, and we find that models initialised with $\bm{B} = +B_0\hat{\bm{z}}$ collapse only slightly faster than their counterparts with $-B_0\hat{\bm{z}}$, but still do not collapse faster than iMHD.  By \rhoapprox{-12}, a weak counter-rotating envelope has formed in \zetam{18} and \zetam{19}, and we have verified that this does not form for the $\bm{B} = +B_0\hat{\bm{z}}$ models.  

\begin{figure}
\begin{center}
\includegraphics[width=\columnwidth]{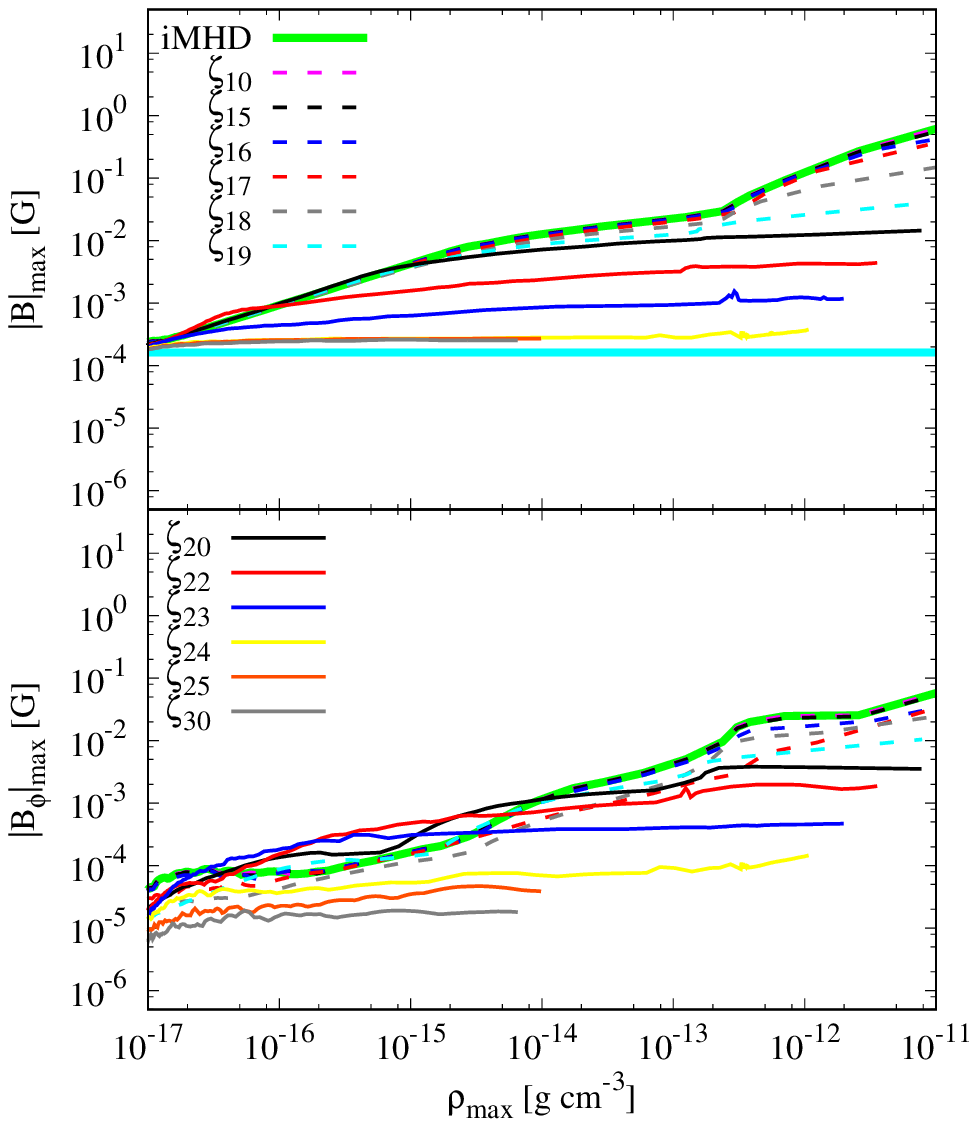}
\caption{The evolution of the maximum magnetic field strength (top) and the magnitude of the maximum toroidal field $|B_\phi |_\text{max}$ (bottom) for selected models.  The thick cyan line represents the initial magnetic field strength.  The legend is split across the panels for clarity.  Initially, the entire magnetic field is poloidal $|B_\text{p}| = \sqrt{B_\text{r}^2 + B_\text{z}^2}$, and the evolution of the poloidal and total magnetic field strengths are similar.  The maximum magnetic field strength increases for all models, although only increases by a factor of \sm1.7 for the low ionisation models.    For the majority of the models, $|B_\text{p}|_\text{max}  > |B_\phi|_\text{max}$, however, \zetam{23}, \zetam{22}, \zetam{20}, \zetam{24}, \zetam{19} and \zetam{18} all switch to having a dominant toroidal field at increasing maximum densities.}
\label{fig:res:bfield}
\end{center}
\end{figure}

Fig.~\ref{fig:res:bfield} shows the maximum magnetic field strength and the magnitude of the maximum toroidal field as a function of maximum density $\rho_\text{max}$, which we use as a proxy for time; recall that $B_{\phi,0} = 0$.
Although the absolute value of the Hall effect is not the strongest in \zetam{22} (i.e. there are larger values of $|\eta_\text{HE}|$ at lower ionisation rates; see Fig.~\ref{fig:ic:nicil}), the net magnetic field in this model is stronger than in \zetam{25} (i.e. the model with the largest $|\eta_\text{HE}|$) since that model has large ambipolar diffusion.  Thus, more of the net magnetic field is converted into the toroidal magnetic field in \zetam{22} than \zetam{25} (or any other model in our suite).  In several models, the maximum toroidal field becomes stronger than the maximum poloidal field at a given location, with $|B_\phi|_\text{max}  > |B_\text{p}|_\text{max}$ occurring for \zetam{23}, \zetam{22}, \zetam{20}, \zetam{24}, \zetam{19} and \zetam{18} at increasing maximum densities.

The maximum magnetic field strength continually increases as the cloud collapses for the models with \zetage{-23}, although the growth rate is much slower for \zetam{22} and \zetam{23}.  By \rhoapprox{-13}, $|B|_\text{max} \approx 5.7B_0$, $19B_0$ and $130B_0$ for \zetam{23}, \zetam{22} and iMHD, respectively;  the low ionisation rate models (i.e. \zetam{24}, \zetam{25} and \zetam{30}) have magnetic field strengths that asymptote at $|B|_\text{max} \approx 1.7B_0$.

\begin{figure}
\begin{center}
\includegraphics[width=\columnwidth]{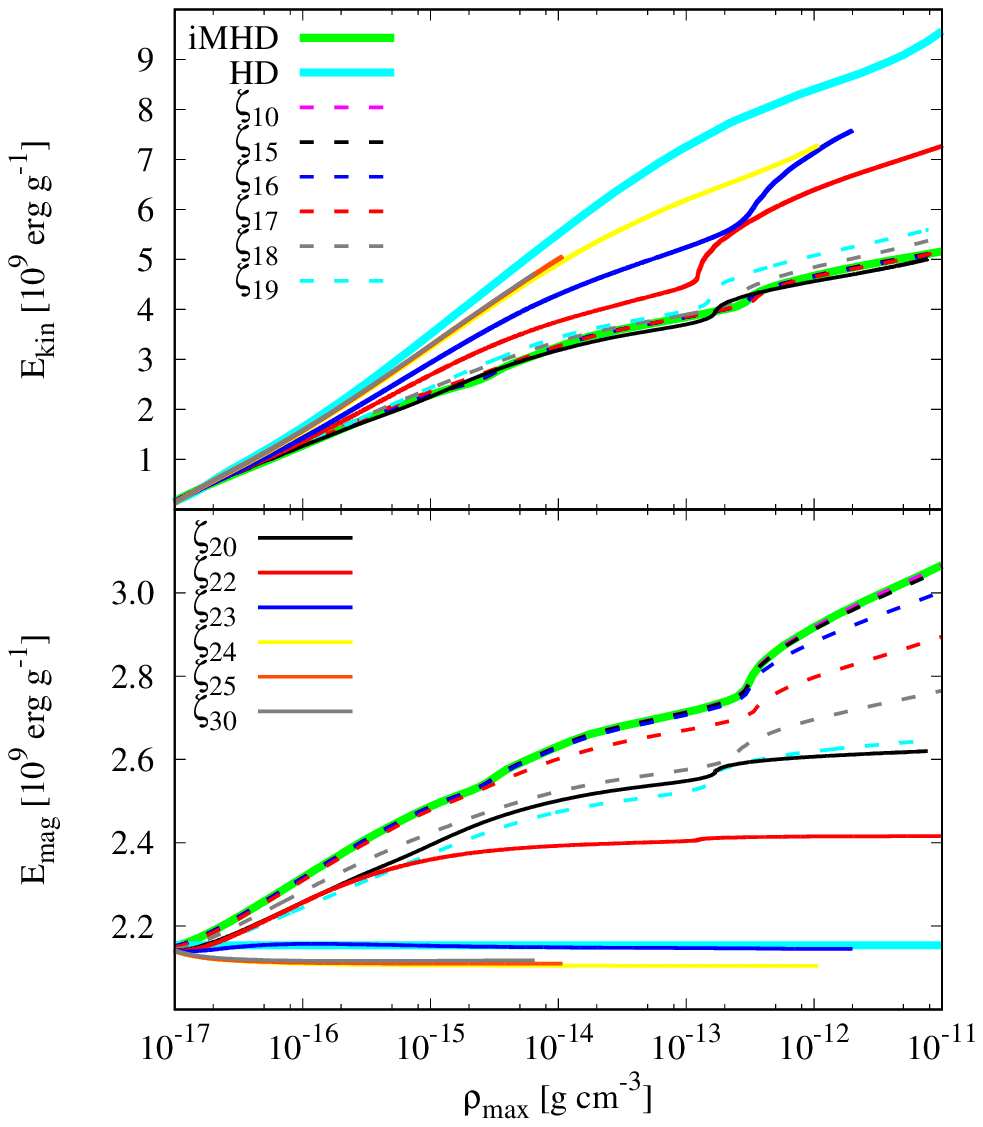}
\caption{Evolution of the total kinetic (top) and magnetic (bottom) energies for selected models.  The thick green and cyan lines represent iMHD and HD, respectively; the thick cyan line in the bottom panel represents the initial magnetic energy, $E_\text{mag,0}$.  The legend is split across the panels for clarity.  When comparing kinetic energy at similar maximum densities, there is typically a progression from the iMHD to HD models as the ionisation rate decreases.  For \zetage{-22}, the total magnetic energy increases as the molecular cloud collapses, whereas it slightly decreases for \zetale{-23}.}
\label{fig:res:trans}
\end{center}
\end{figure}

Fig.~\ref{fig:res:trans} shows the evolution of the total kinetic and magnetic energies for selected models.  These values include the gas in both the cloud and background medium.
Magnetic fields support the molecular cloud against collapse, thus at any given \rhomax, there is more kinetic energy in HD than in iMHD, with the kinetic energy decreasing from the HD value to the iMHD value as \zetacr \ is increased; this can also be seen in Figs.~\ref{fig:res:slice:vr} and \ref{fig:res:slice:vy}, which show decreasing radial and azimuthal velocities for increasing \zetacr.  The total magnetic energy increases for models with \zetage{-22}, with the magnetic energy growing more slowly for models with lower \zetacr.  The total magnetic energy decreases for the models with \zetale{-23}; at the end of the simulation, the models with \zetale{-24} have decreased by \sm2 per cent, while \zetam{23} has decreased by only \sm0.4 per cent.

\subsection{Approaching ideal MHD}
\label{sec:res:ideal}
As ionisation rates increase, the medium becomes more ionised, thus the ionisation fractions begin to approach the ideal MHD limit; however, at temperatures and densities presented here, ionisation fraction remains $\ll 1$.  The question we ask in this section is how high of an ionisation rate is required to safely approximate the ideal MHD limit?  

From the induction equation (Eqn.~\ref{eq:ind}), a one-fluid non-ideal MHD system can be approximated as ideal when $\left.\text{d} \bm{B}/\text{d} t\right|_\text{non-ideal} \ll  \left(\bm{B} \cdot \nabla\right) \bm{v} - \bm{B} \left(\nabla\cdot \bm{v}\right)$.  However, we want to know the largest possible $\left.\text{d} \bm{B}/\text{d} t\right|_\text{non-ideal}$ that will still result in a system equivalent to iMHD.  Since different quantities may show equivalence at different \zetacr, we will compare several different quantities below, including 2D cross sections, total and maximum values.   A non-ideal MHD model must be equivalent in all of our quantities to be deemed equivalent to iMHD.

\begin{figure}
\begin{center}
\includegraphics[width=\columnwidth]{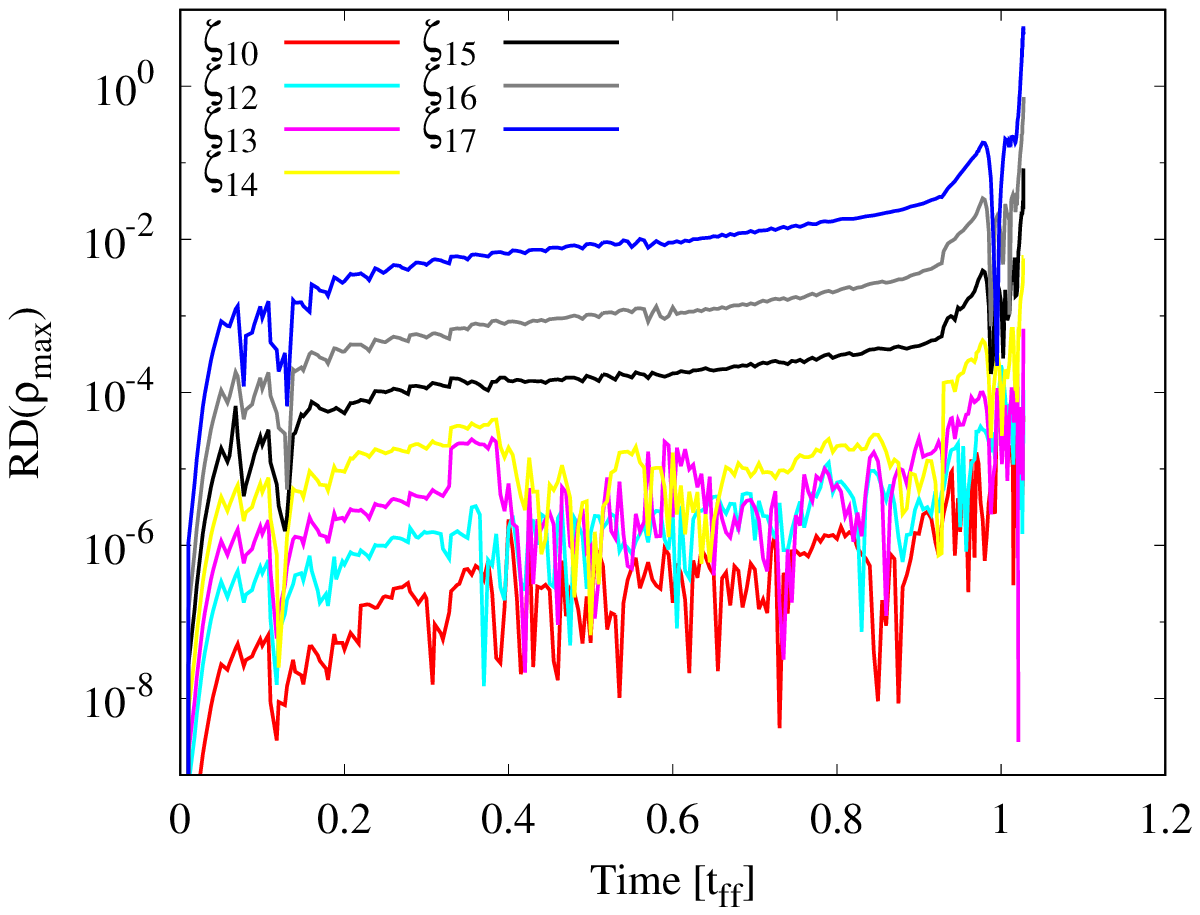}
\caption{The relative difference, as defined in \eqref{eq:rd}, of the maximum density with respect to time for the high ionisation rate models compared to iMHD.  The relative differences are less than 10 per cent for $t < $ \tff, however, even slight differences in evolution times at high densities (i.e. at $t \sim$ \tff) cause large relative differences.  The evolution times of the models with \zetage{-14} agree within $10^{-4}$} for the majority of the collapse.
\label{fig:res:rd:high:rho}
\end{center}
\end{figure}

\begin{figure}
\begin{center}
\includegraphics[width=\columnwidth]{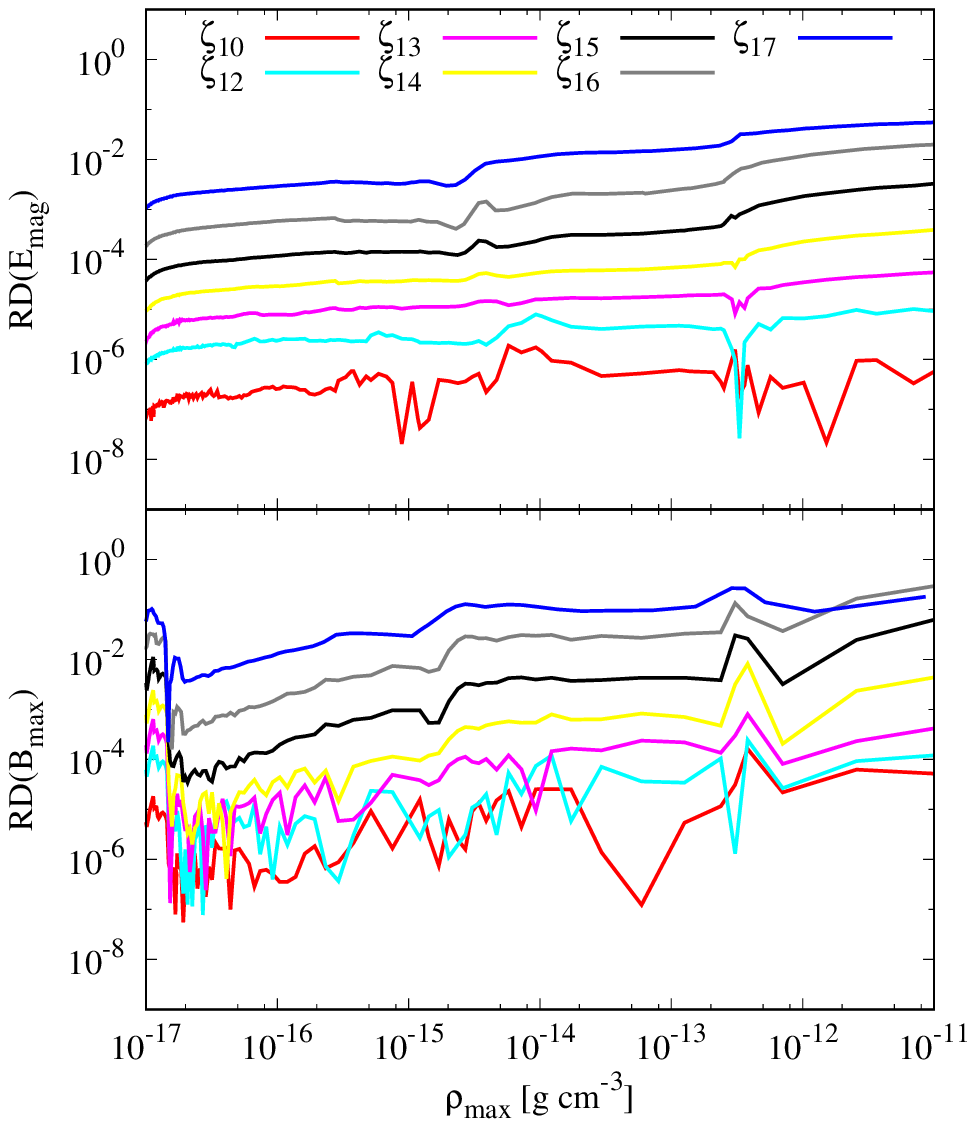}
\caption{The relative differences of the magnetic energy (top) and the maximum magnetic field strength (bottom) of the high ionisation rate models compared to iMHD.  For \rhole{-11}, the magnetic energy of the high ionisation rate models differ by less than 10 per cent from iMHD, and the maximum magnetic field strength differs by less than 40 per cent.}
\label{fig:res:rd:high}
\end{center}
\end{figure}

Upon visual inspection of the cross section plots in Figs.~\ref{fig:res:slice:rho} -- \ref{fig:res:slice:vy}, we find that the models with \zetage{-15} look similar to iMHD for all properties.  This also appears true for the evolution of the maximum density with time (Fig.~\ref{fig:res:rhoVt}) and for total kinetic and magnetic energy (Fig.~\ref{fig:res:trans}).  To quantify this,  Fig.~\ref{fig:res:rd:high:rho} shows the relative difference,
\begin{equation}
\label{eq:rd}
RD \equiv \frac{\left| X\left(\zeta_n\right) - X\left(\text{iMHD}\right)\right|}{\left|X\left(\text{iMHD}\right)\right|},
\end{equation}
where $X\left(\zeta_n\right)$ and $X\left(\text{iMHD}\right)$ are the values for the non-ideal model $\zeta_n$ and iMHD, respectively, of the maximum density, \rhomax, and Fig.~\ref{fig:res:rd:high} shows the relative difference of magnetic energy and total magnetic field strength of a high ionisation rate model compared to the iMHD model; for comparison, we intentionally include models that clearly do not approach the ideal limit. 


The maximum density in the models with \zetage{-15} differs from iMHD by less than 10 per cent for the entire calculation, with only \zetam{15} yielding a difference larger than one per cent near the end.

The relative difference of the magnetic energy and maximum magnetic field strength compared to $\rho_\text{max}$ decreases for increasing ionisation rates; models with higher ionisation rates are more similar to iMHD.  The magnetic energy of all models with \zetage{-17} differs from iMHD by less than 10 per cent, while the energy in models with \zetage{-15} differ by less than one per cent.

Total magnetic energy is a global property of the simulation, including both the quickly collapsing inner region and the slowly evolving background medium.  The maximum magnetic field strength, however, is localised in or near the core, and is sensitive to the evolution of the collapse and is obtained from a single particle; thus, we do not expect as low of relative difference as in the energies.  Models with \zetage{-12} have a maximum magnetic field strengths that differ from iMHD by less than $10^{-4}$, while models with \zetage{-14} differ by less than one per cent.  The relative difference between \zetam{15} and iMHD increases to almost 10 per cent in the first hydrostatic core; this large difference and the increasingly large difference in RD(\rhomax) suggests that \zetam{15} does not approach the ideal MHD limit.

For \zetage{-14}, the grain properties have minimal effect on the non-ideal MHD coefficients (see Section~\ref{sec:IC:grain}), except when decreasing to smaller grains of uniform size.  Although our conclusions will be qualitatively unaffected by switching to larger grains or to an MRN grain distribution, the agreement between (e.g.) \zetam{14} and iMHD may not be as robust for the smaller grain size of \graina{0.01}.

For the collapse up to \rhoeq{-11}, we conclude that non-ideal MHD models with \zetage{-14} agree with ideal MHD within one per cent, and models with \zetage{-13} agree with ideal MHD within of 0.1 per cent, suggesting that they are essentially indistinguishable from ideal MHD models.

\subsection{Approaching pure hydrodynamics}
\label{sec:res:hydro}
As ionisation rates decrease, the medium becomes more neutral, thus begins to approach the purely hydrodynamic limit.  Unlike approaching the ideal limit where all simulations include magnetic fields and non-ideal effects begin to negligibly contribute to their evolution, the hydrodynamic models by definition exclude magnetic fields; thus, for a low-ionisation rate model to approach the HD model, their evolution must essentially ignore the magnetic field.  

In order that the one-fluid non-ideal MHD equations reduce to the ideal MHD limit, all that has to happen is for the third term in Eqn.~\ref{eq:ind} to become negligible.  Thus, it is unsurprising that with a high enough ionisation rate, the ideal MHD limit is recovered to a high level of accuracy (see Section~\ref{sec:res:ideal}).  To recover the hydrodynamical limit, however, all four terms in the induction equation (Eqn.~\ref{eq:ind}), which are calculated using the magnetic field and density of a particle and its neighbours must sum to exactly zero, and the second term in the momentum equation (Eqn. \ref{eq:mom}) must also be zero. Thus, we do not expect to be able to approximate the hydrodynamic limit using non-ideal MHD to the same level of accuracy that was achieved for the ideal limit.

As can be seen in the cross section plots (Figs.~\ref{fig:res:slice:rho} -- \ref{fig:res:slice:vy}), \zetam{24} shares more characteristics with HD than iMHD or even \zetam{23}; however, there are still noticeable structure differences.  Model HD has a slightly larger scale height, and slightly faster rotational and radial velocities.  Based upon visual inspectional alone, we conclude that by decreasing the ionisation rate, the models approach HD, however,  \zetam{24} is not equivalent to HD.  At \rhoapprox{-15}, \zetam{24}, \zetam{25} and \zetam{30} are indistinguishable from one another, although, neither of the latter two models progressed to \rhoapprox{-13} for a later comparison.

\begin{figure}
\begin{center}
\includegraphics[width=\columnwidth]{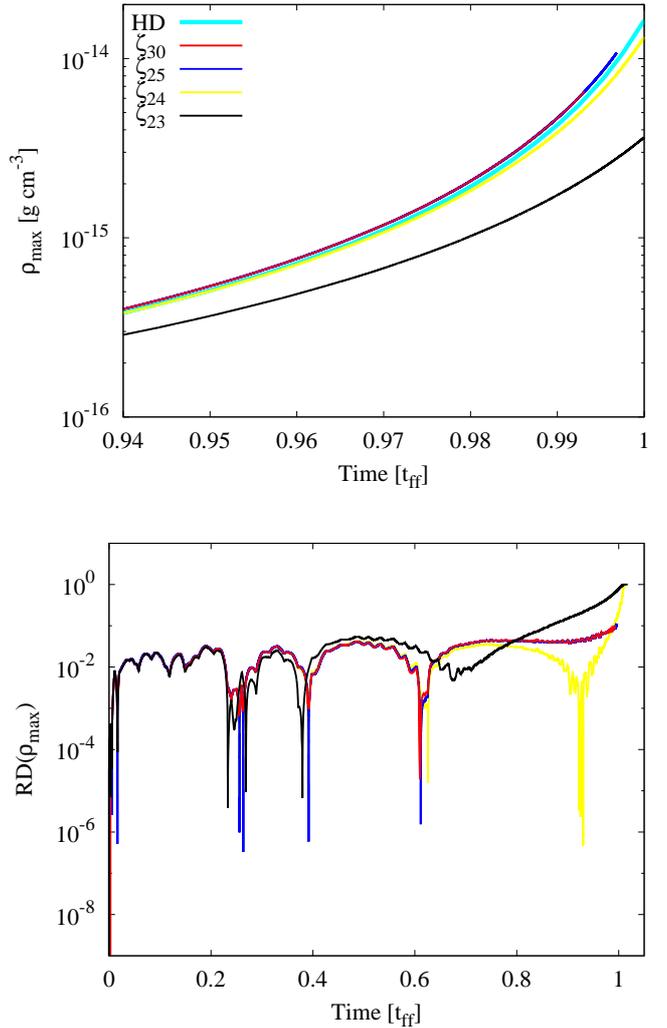}
\caption{\emph{Top}: Evolution of the maximum density over a short range of time for our low ionisation rate models, as in Fig.~\ref{fig:res:rhoVt}.  \emph{Bottom}: The relative difference of the maximum density with respect to time over the entire evolution, as in Fig.~\ref{fig:res:rd:high:rho}.  The collapse rates are slightly faster for \zetam{25} and \zetam{30} compared to HD, with the relative difference never surpassing 10 per cent.  Model \zetam{24} collapses slightly slower than HD, and the relative difference surpasses 10 per cent at $t \approx 0.99$\tff.}
\label{fig:res:rd:low:rho}
\end{center}
\end{figure}

For our quantitative analysis, Fig.~\ref{fig:res:rd:low:rho} shows \rhomax \ and the relative difference in \rhomax \ of the low ionisation rate models compared HD.
The evolution of \zetam{23} has been included for reference, but it clearly cannot approximate HD.  The collapse rate for these low ionisation rate models varies in time with respect to HD, collapsing slightly faster or slower depending on the time; the transitions are marked by the relative difference approaching zero in these plots.  Over the entire evolution, \zetam{24} collapses slower than HD, with the relative difference surpassing 10 per cent by $t \approx 0.99$\tff.  Given the slower collapse and the visual differences, \zetam{24} is not equivalent to HD.  Models \zetam{25} and \zetam{30} collapse faster than HD but evolve at the same rate as one another, and at any given time, their maximum density differs from HD by less than 10 per cent.

In \zetam{24}, \zetam{25} and \zetam{30}, the maximum magnetic field strength grows by a factor of \sm1.7 (bottom panel of Fig.~\ref{fig:res:bfield}); for reference, by this density of \rhotwoapprox{5}{-15}, the magnetic field strength in iMHD has increased by a factor of \sm60.  Despite the maximum magnetic field strength slightly increasing, the total magnetic energy in these three models decreases to $E_\text{mag} \approx 0.98E_\text{mag,0}$ (bottom panel of Fig.~\ref{fig:res:trans}).

If the grain size was decreased to \graina{0.01} or an MRN distribution was used, then the non-ideal MHD coefficients would be lower thus these models would be slightly more ideal (Fig.~\ref{fig:ic:nicil:grains}).  Therefore, modelling smaller grains would yield low ionisation rate models that are even less similar to the hydrodynamical case than those presented here.  If an \graina{1} grain size was modelled, then the non-ideal MHD coefficients would be larger, suggesting these models would be more similar to HD.  However, the larger coefficients results in a smaller timestep (recall Eqn.~\ref{eq:dt:ni}), which makes it prohibitively expensive to model; see further discussion in Section~\ref{sec:timing}.

Thus, even with low cosmic ray ionisation rates of \zetale{-24}, the relative differences of the energies, magnetic field strengths and the maximum densities remain between one and 10 per cent.  However, given the slow growth of the magnetic energy and maximum magnetic field strength, we can conclude that models with \zetale{-24} approach the hydrodynamic model.  At these low ionisation rate, unless precision results are required, pure hydrodynamics can be used in place of non-ideal MHD.

\subsection{Timestepping}
\label{sec:timing}

Given unlimited resources and time, one should simply use non-ideal MHD with the desired cosmic ray ionisation rate rather than ideal or hydrodynamic approximations.  However, including non-ideal MHD adds to the computational expense both by additional calculations and by decreasing the minimum timestep.  

\begin{figure}
\begin{center}
\includegraphics[width=\columnwidth]{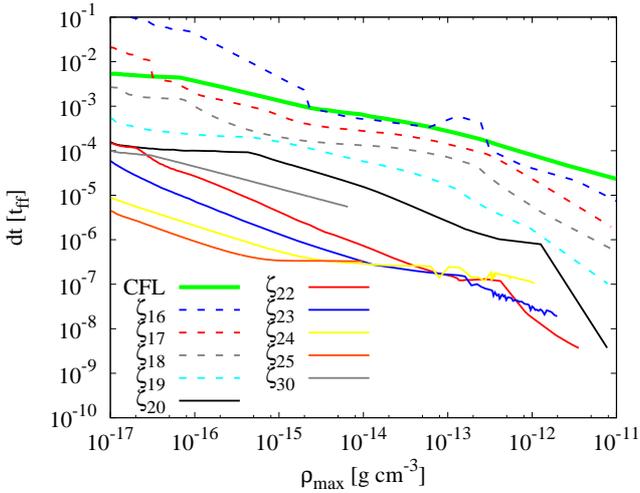}
\caption{The evolution of the minimum timestep as calculated from the non-ideal terms (see Eqn.~\ref{eq:dt:ni}); the thick green line is the timestep calculated from the Courant-Friedrichs-Lewy (CFL; see Eqn.~\ref{eq:dt:cfl}) condition, and is included for reference.  For the models whose timesteps are controlled by Ohmic resistivity or ambipolar diffusion, we plot d$t = \text{d}t_\text{OR,AD}/N_\text{sts}$, where $N_\text{sts}$ is given in \eqref{eq:nsts}.  Initially, \zetam{25} requires the smallest timesteps, while \zetam{30} uses a timestep $\approx20-50$ times larger due to super-timestepping with the Ohmic time constraint.  Models with \zetage{-15} have d$t_\text{non-ideal} > \text{d}t_\text{CFL}$ thus are excluded.}
\label{fig:dis:dt:pred}
\end{center}
\end{figure}

Fig.~\ref{fig:dis:dt:pred} shows the minimum and CFL timesteps at each density.
All model with \zetage{-15} have d$t_\text{non-ideal} > \text{d}t_\text{CFL}$, while \zetam{17} has d$t_\text{non-ideal} \sim \text{d}t_\text{CFL}$.  Models with $10^{-25} \lesssim \zeta_\text{cr}/\text{s}^{-1} \lesssim 10^{-19}$ are always limited by the Hall timestep, while \zetam{17} and \zetam{18} are Hall-limited for \rhotwoge{2}{-15} and \rhoge{-16}, respectively.  Model \zetam{30} is limited by the super-timestepping algorithm using the Ohmic timestep.  Note that the ambipolar or Ohmic timesteps would be the limiting case in most models if super-timestepping were not used.  

At \rhoapprox{-15}, the non-ideal timestep in \zetam{25} is limited by the Hall timestep and is \sm3900 times shorter than the CFL timestep.  At \rhoapprox{-11}, the non-ideal timestep in \zetam{17} (which uses the fiducial cosmic ionisation ray rate) is \sm15 times shorter than the CFL timestep.  Given that studies of collapsing molecular clouds need to reach maximum densities at least a few orders of magnitude larger than presented here, the slow-down of a factor of a few for the canonical ionisation rate of \zetaapprox{-17} can be tedious, while it can be completely prohibitive for lower ionisation rates.

\begin{figure}
\begin{center}
\includegraphics[width=\columnwidth]{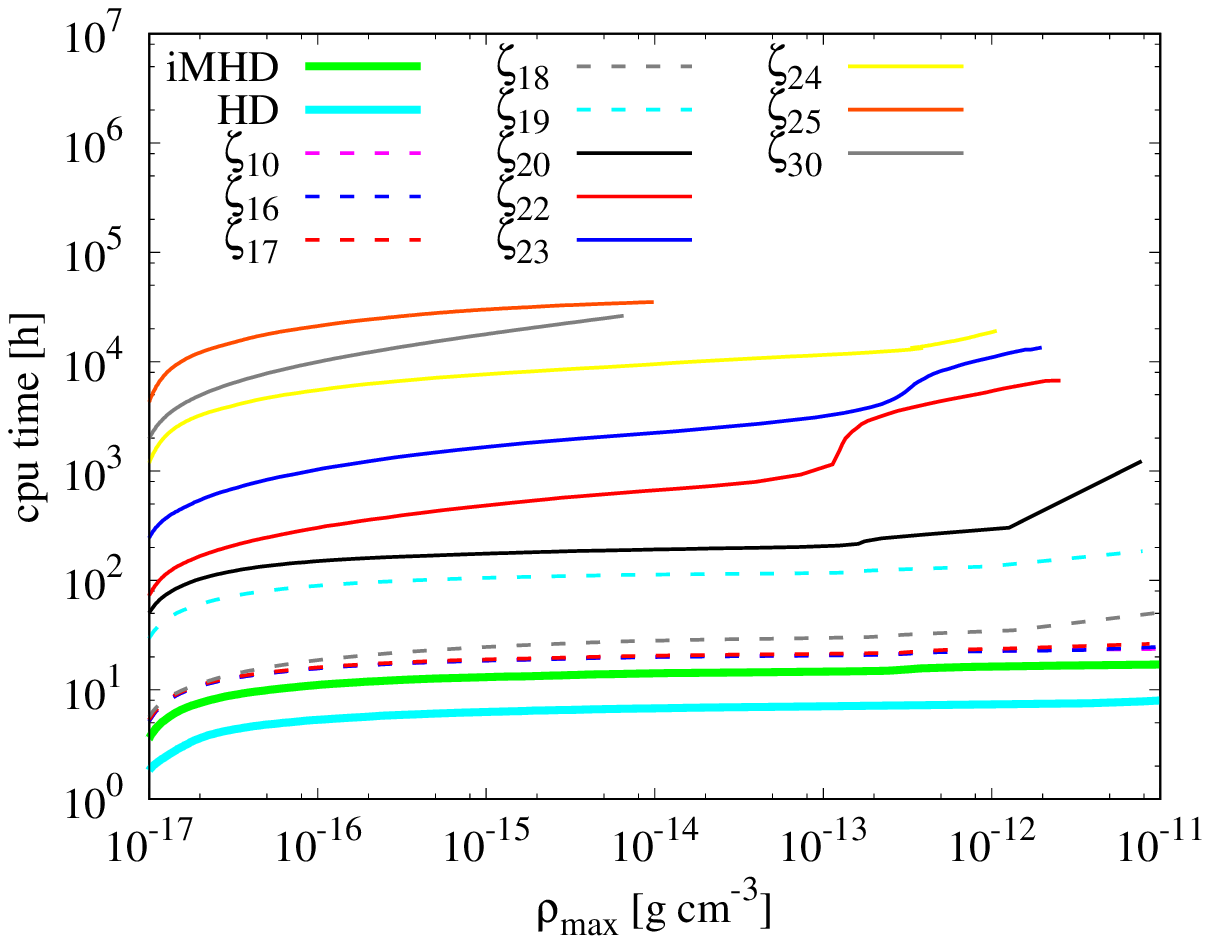}
\caption{The cumulative CPU time used for each model.  HD, iMHD and models with \zetage{-18} run very quickly to \rhoeq{-11}.  As the ionisation rate decreases, the required resources becomes prohibitively expensive, especially for \zetale{-24}.  To reach \rhoapprox{-15}, it takes \zetam{25} \sm2200 times longer than iMHD.  All the non-ideal models with \zetage{-17} take the same length of time, which is \sm1.4  times longer than iMHD to reach \rhoapprox{-11}.}
\label{fig:dis:tcpu}
\end{center}
\end{figure}

The effect of non-ideal MHD on performance is very evident in the number of CPU hours used, as shown in Fig.~\ref{fig:dis:tcpu}.
The HD model runs \sm53 per cent faster than the iMHD model in part due to the reduce number of calculations required and the slightly longer CFL timestep since $v_\text{A}\equiv0$.  The models with \zetage{-17} take \sm1.4 times longer to reach \rhoapprox{-11} than iMHD; since their non-ideal timesteps are similar to the CFL timestep, the additional time used is mostly taken up by the \textsc{Nicil} library.  Given that the models with \zetage{-14} well approximate the ideal MHD model (see Section~\ref{sec:res:ideal}), an ideal MHD model can instead be run at a speed-up of \sm1.4.

As the ionisation rate decreases, the cumulative CPU time increases, with \zetam{25} being the most expensive simulation followed by \zetam{30} and then the remainder of the models in order of increasing ionisation rate.  To reach \rhoapprox{-15}, it takes \zetam{25} \sm2200 times longer than iMHD, thus these models are prohibitively expensive to run to any useful maximum density.  However, since the models with \zetale{-24} have small relative differences for their total magnetic energies and  maximum magnetic field strengths, respectively, it would be reasonable to run purely hydrodynamical models in their place.

\section{Summary and conclusion}
\label{sec:conclusion}

We have presented a suite of non-ideal magnetohydrodynamics simulations with various cosmic ray ionisation rates \zetacr \ to determine what rate is required to recover a hydrodynamical collapse and an ideal MHD collapse.  Our models were initialised as a 1 M$_\odot$, spherically symmetric, rotating molecular cloud core; the cloud was magnetised with a magnetic field initially aligned anti-parallel to the rotation axis, and had an initial strength of $B_0 = 1.63 \times 10^{-4}$ G, or $\mu_0=5$.  Our models used a uniform grain size of \graina{0.1}, but we discussed how different grain models are expected to change the results.  In particular, we found that at high cosmic ray ionisation rates the results will be approximately independent of grain properties, but at very low ionisation rates the non-ideal MHD coefficients that we uses tend to be larger than those produced by MRN grain size distributions, suggesting that the ionisation rates required to approximate hydrodynamical evolution may be even lower than the values we find using \graina{0.1}.  All three non-ideal MHD terms (Ohmic resistivity, Hall effect and ambipolar diffusion) were included.  We tested 15 different cosmic ray ionisation rates, which were held constant for the entire simulation.  The initial density of the molecular cloud core was $\rho_0=7.43\times 10^{-18}$ \gpercc, and it was evolved until $10^{-15} \lesssim \rho_\text{max}/($\gpercc$) \lesssim 10^{-11}$, depending on the value of \zetacr.  

Our two key results are as follows:
\begin{enumerate}
\item \emph{Approaching the ideal MHD limit}:   We evolved models with high ionisation rates until they entered the first hydrostatic core and reached \rhoapprox{-11}.  Models with \zetage{-13} were indistinguishable from the ideal MHD model when considering the evolution of their maximum density, magnetic energy and magnetic field strengths i.e. all properties matched the evolution of the ideal MHD model within 0.1 per cent.  The evolution of the model with \zetaeq{-14} matched the evolution of the ideal MHD model within 1 per cent.

\item \emph{Approaching the hydrodynamic limit}:  The models with \zetale{-24} look similar to the hydrodynamical model, but with noticeable differences; in these models, the total magnetic energy grew by a factor of 1.7 and the maximum magnetic field strength decreased by 2 per cent.   Our lowest ionisation rate models were not followed beyond the isothermal collapse phase due to the non-ideal MHD constraints on the timestep.  Those with \zetale{-25} had maximum densities at a given time that agreed with the hydrodynamical model within 10 per cent.  Given the reasonable agreement with the hydrodynamical model and the orders of magnitude increase in runtime, we conclude that hydrodynamical models can be used to approximate non-ideal MHD models with \zetale{-24} unless precision results are required.
\end{enumerate}

We conclude that it is possible to reproduce ideal MHD and purely hydrodynamical collapses using non-ideal MHD given an appropriate cosmic ray ionisation rate.  However, reaching either limit by cosmic ray ionisation alone is unlikely, since molecular clouds in the local neighbourhood have cosmic ray ionisation rates of \zetaapprox{-17} \citep[e.g.][]{PadovanGalliGlassgold2009,NeufeldWolfire2017}.  Even if the comic ray ionisation rate were to temporarily increase (e.g. due to a nearby supernova), the increase will not be sustained enough to shift the evolution into the ideal MHD regime for the lifetime of the initial collapse.  

For cosmic ray ionisation rates of \zetale{-24}, we can approximate a hydrodynamic collapse, however, it is improbable to reach these low ionisation rates in local diffuse molecular clouds given that this rate is well below the rate expected from radionuclide decay.  Studies of the early universe by \citet{SusaDoiOmukai2015} found that, even in the absence of cosmic rays and heavy metals, the primordial gas is partially ionised (i.e. contains e$^-$, H$^+$, Li$^+$).  Their electron fractions for this case are similar to ours for \zetale{-24}, however, their H$^+$, Li$^+$ fractions are higher due to the absence of dust grains.  Although our low ionisation rate models are unlikely to be relevant in the local Universe, they may have implications for the early Universe.

While our quantitative conclusions depend on our initial conditions and our chosen maximum densities, we find that the canonical cosmic ray ionisation rate of \zetaeq{-17} approaches neither the ideal MHD limit nor the hydrodynamical limit. For star formation simulations to properly evolve the magnetic field, non-ideal MHD is essential.

\section*{Acknowledgements}

JW and MRB acknowledge support from the European Research Council under the European Community's Seventh Framework Programme (FP7/2007- 2013 grant agreement no. 339248).  DJP received funding via Australian Research Council grants FT130100034, DP130102078 and DP180104235.  This work was supported by resources on the swinSTAR national facility at Swinburne University of Technology. swinSTAR is funded by Swinburne and the Australian Government's Education Investment Fund. 

\bibliography{ExtremeIonisation.bib}

\label{lastpage}
\end{document}